\documentclass[fleqn,usenatbib]{mnras}

\usepackage{newtxtext,newtxmath}

\usepackage[T1]{fontenc}

\DeclareRobustCommand{\VAN}[3]{#2}
\let\VANthebibliography\thebibliography
\def\thebibliography{\DeclareRobustCommand{\VAN}[3]{##3}\VANthebibliography}

\usepackage{graphicx}	
\usepackage{amsmath}	
\usepackage{subfig}

\title[XTE J2012+381]{Black hole spin estimation of XTE J2012+381 using simultaneous observations of \emph{Swift/XRT} and \emph{NuSTAR}}

\author[Raj Kumar]{
Raj Kumar,$^{1,2}$\thanks{E-mail: arya95raj@gmail.com}
\\
{$^1$Astrophysical Sciences Division, Bhabha Atomic Research Centre, Mumbai - 400 085, India}\\
{$^2$Homi Bhabha National Institute, Mumbai - 400 094, India} \\}

\date{Accepted XXX. Received YYY; in original form ZZZ}

\pubyear{2023}

\begin{document}
\label{firstpage}
\pagerange{\pageref{firstpage}--\pageref{lastpage}}
\maketitle

\begin{abstract}
A sufficiently precise measurement of black hole spin is required to carry out quantitative tests of the Kerr metric and to understand several phenomena related to astrophysical black holes. After 24 years, XTE J2012+381 again underwent an outburst in December 2022. In this work, we focused on the measurement of spin and mass of black hole candidate XTE J2012+381 using broadband spectral analysis of X-ray data from \emph{Swift/XRT} and \emph{NuSTAR}. By using the relxillCp model, the spin and inclination of the source were found to be $0.883_{-0.061}^{+0.033}$ and $46.2_{-2.0}^{+3.7}$ degrees, respectively for high disk density ($i.e.\;10^{20}\;\mathrm{cm}^{-3}$). We further test our results for lamp post geometry using the relxilllpCp model. The spin and inclination of the source were found to be $0.892_{-0.044}^{+0.020}$ and $43.1_{-1.2}^{+1.4}$ degrees, respectively. Then "continuum-fitting" method was used for the soft state to estimate the mass of BH and found to be $7.95_{-3.25}^{+7.65}\;\mathrm{M}_{\odot}$ and $7.48_{-2.75}^{+5.80}\;\mathrm{M}_{\odot}$ for the spin and inclination estimated from the relxillCp and relxilllpCp model respectively. We used a distance of 5.4 kpc as measured by Gaia using the parallax method. This study also addresses the issue of supersolar iron abundance in XTE J2012+381 by using \texttt{reflionx} based reflection model and found high disk density for the source.
\end{abstract}




\section{Introduction}

Astrophysical black holes (BHs) are mainly characterized by their mass $M$ and spin $J$. The condition of existence of the event horizon, i.e., $J/M^2 < 1$, constrains the spin parameter of BHs. The spin parameter of BHs can be estimated by the gas accretion process for X-ray binaries (XRBs). The spin alters the gravitational field near the event horizon of black holes. There are two leading methods to measure the black hole spin by studying the X-ray radiation emitted from the inner part of the accretion disk. One is the continuum-fitting (CF) method that models the thermal emission from the accretion disk \citep{Zhang1997}. \cite{NT1973} described the relativistic generalization of accretion disk model developed by \cite{SS1973}. They described the emission from a geometrically thin and optically thick disk at each radius of the disk as a function of the mass of black hole $M$, mass accretion rate $\dot{M}$ and spin parameter $a$. Ample observational evidence exists to support that disk sharply truncates at the innermost stable circular orbit (ISCO) during the soft state \citep{Gierlinski2004, Abe2005, McClintock2009, Dunn2010}. Theoretical studies showed that simulated disks do not differ by a large amount from \cite{NT1973} model \citep{Shafee2008, Reynolds2008, Penna2010}. So, for a given value of $M$, one can estimate the spin parameter of BH during the soft state. $D$ and $i$ are required to measure the $R_{ISCO}$. This way, the thermal continuum fitting method can be used to measure the spin parameter of BH for known values of $D$, $i$ and $M$. \citet{McClintock2014} used the ``continuum fitting" method to determine the spin parameter of ten stellar mass BHs.

Another method that is based on the study of the disc's reflection spectrum is ``relativistic reflection" (or iron line method) \citep{Febian1989, Tanaka1995}.  Thermal photons emitted from the inner region of the accretion disk are comptonized by the hot electrons in the corona. A fraction of comptonized photons is further absorbed and reemitted by the accretion disk to produce the reflection spectrum. The reflection spectrum usually contains a strong feature of the Iron $\mathrm{k}\alpha$ line, absorption edge, and Compton hump. The observed Iron $\mathrm{k}\alpha$ line is very broad and skewed due to relativistic effects occurring near the strong gravity of Kerr BH \citep{Febian2000, Reynolds2003, Miller2007}. The reflection spectrum of Kerr BHs depends upon various disk parameters and emissivity profiles of the accretion disk that depend upon disk geometry. The "relativistic reflection" method does not require M and D as the continuum-fitting method requires them. \citet{Draghis2023a} estimated the spin of ten stellar-mass black holes using the "relativistic reflection" method. These two methods are independent and useful for estimating the black hole spin. The CF method is applicable only to stellar-mass black holes. Whereas X-ray reflection spectroscopy can be applied to stellar masses and supermassive black holes. Recently, \citet{Reynolds2021} discussed these BH spin measurement techniques in detail.

Except for a few cases, an independent spin measurement from these two methods was found to be within good agreement. Broadband spectroscopy can be used to constrain different spectral parameters like $M$, $D$, $a$ and $i$ in a narrow region of parameter space. This can be done by combining these two methods. \citet{Parker2016} used the constraints of BH spin and inclination from the reflection method as input to continuum fitting to estimate the mass and distance of BHXRB GX 339-4. Few other studies also used both methods simultaneously to constrain the different physical parameters of BHXRBs i.e.  MAXI J1348-630 \citep{Kumar2022} GRS 1716–249 \citep{Tao2019}, LMC X–3 \citep{Jana2021}, and GX 339–4 \citep{Parker2016}.

In this work, we focused on the estimation of the spin, mass of the black hole and inclination of the accretion disk using broadband spectral analysis. XTE J2012+381 is a black hole candidate \citep{White1998} discovered by \emph{RXTE} all-sky monitor during an outburst in 1998 \citep{Remillard1998}. \cite{Hynes1999} confirmed an optical counterpart of BH. \cite{Campana2002} put a limit of $3.7 d_{10}\mathrm{M}_\odot$ and $22 d_{10}\mathrm{M}_\odot $ for no spin and maximally spinning BH, respectively, using spectral analysis of \emph{BeppoSAX} observations. The measured distance of XTE J2012+381 is $5.4_{-1.50}^{+3.39}$ kpc using parallax method \cite{Gaia2016}. Recently, in December 2022, it underwent an outburst that was detected by \emph{MAXI} telescope \citep{Kawamuro2022} and was further confirmed by \emph{Swift/XRT} \citep{Kennea2022}. When \emph{NICER} started to observe the source, it had already reached the intermediate state. \cite{Draghis2023} analyzed the 2 \emph{NuSTAR} and 105 \emph{NICER} observations. They measured the spin of BH and the inclination of the accretion disk $0.988_{-0.030}^{+0.008}$ and $68_{-11}^{+6}$ degrees, respectively. 

The paper is comprised of 4 sections. In Section 2, we provide the details of the observed data and the data reduction. Section 3 is dedicated to the spectral modelling of \emph{Swift/XRT} and \emph{NuSTAR} data. The results are discussed and concluded in Section 4.

\section{Observations and data reduction.}

 \subsection{Swift/XRT}
\begin{figure}
\centering
\includegraphics[height=0.5\textwidth, angle=-90]{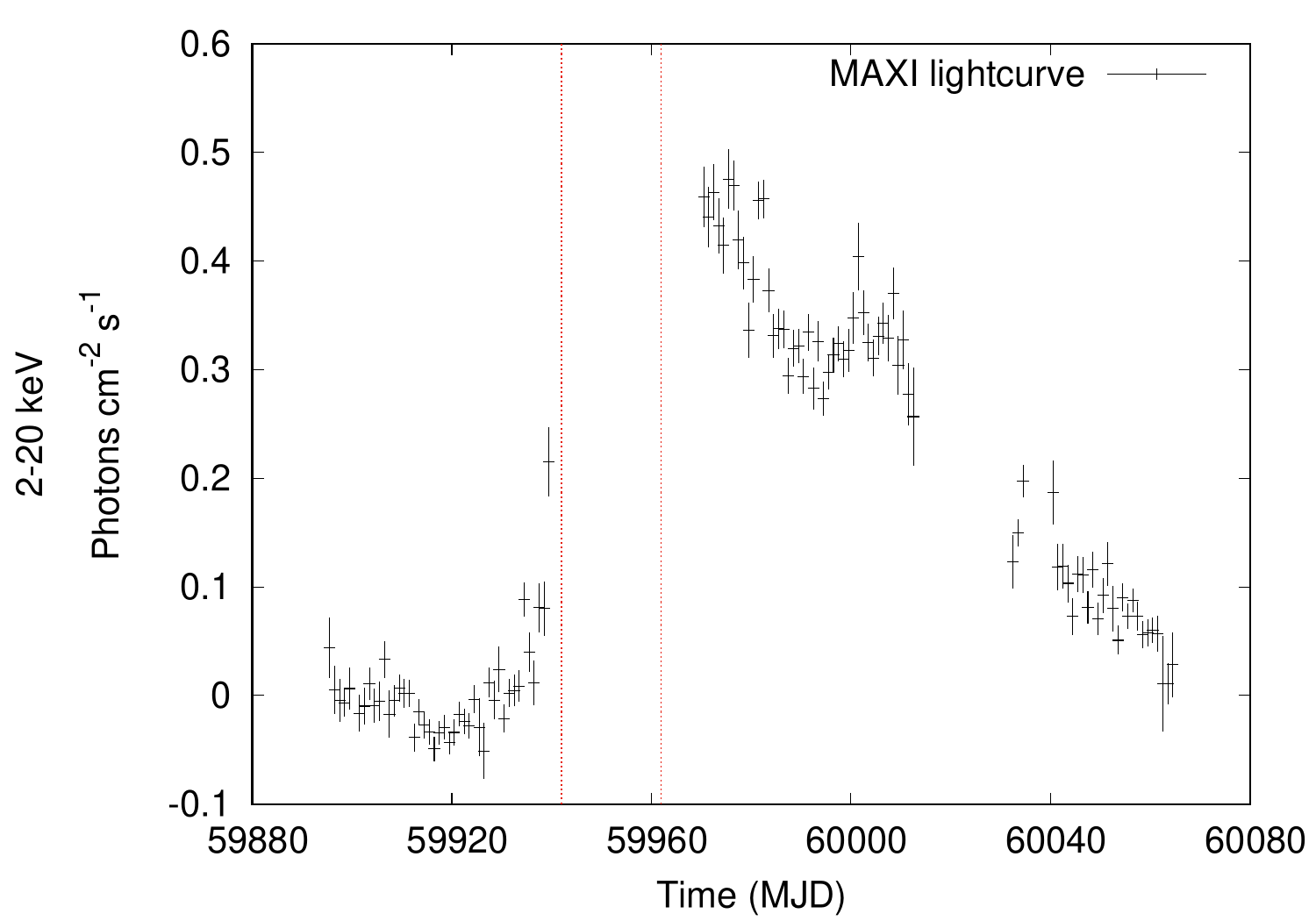}
\caption{The MAXI lightcurve of XTE J2012+381 in the 2–20 keV energy range as observed by MAXI/GSC during the 2022-2023 outburst. The vertical lines (red color) correspond to Swift/XRT and NuSATR observations used in this work.}
\label{lc1}
\end{figure}

 \emph{Swift} and \emph{NuSTAR} observed the XTE J2012+381 simultaneously on MJD 59942 (2022, December 29) and MJD 59962 (2023, January 18). The details of observations are given in Table \ref{data} and MAXI lightcurve is shown in Figure \ref{lc1}. These observations belong to the soft state and are suitable for spin measurement. The X-ray telescope \emph{XRT} \citep{Burrows2005} is the X-ray instrument of the Swift Observatory, working in 0.2-10 keV energy range \citep{Gehrels2004}. We analyzed the \emph{Swift/XRT} obs ID 00089581002 and obs ID 00015473007 available in WT mode. We used \texttt{HEASOFT-6.31.1} and \texttt{xrtpipeline v.0.13.7} for \texttt{Swift/XRT} data reduction. We take an annular source region with inner and outer radii of 3 and 30 pixels, respectively, for obs ID 00015473007 due to the count rate being greater than 100 $\mathrm{counts\;s^{-1}}$ \citep{Romano2006}. For obs ID 00089581002, a circular source region with 30-pixel radii was chosen. The background region of 30-pixel radii was chosen away from the source region shifted along the window. The $n_H$ value for XTE J2012+381 is $1.28\times10^{22}\;\mathrm{cm^{-2}}$ \citep{White1998}. So, we used grade 0 only to extract the spectrum. The latest RMF provided by the POC team was used, and arf was generated using \texttt{XRTMKARF} task. Further, we added $3\%$ systematic in spectra and rebin for minimum 30 counts per bin using \texttt{GRPPHA} task. 
\subsection{NuSTAR}
\begin{figure}
\centering
\includegraphics[height=0.42\textwidth, angle=-90]{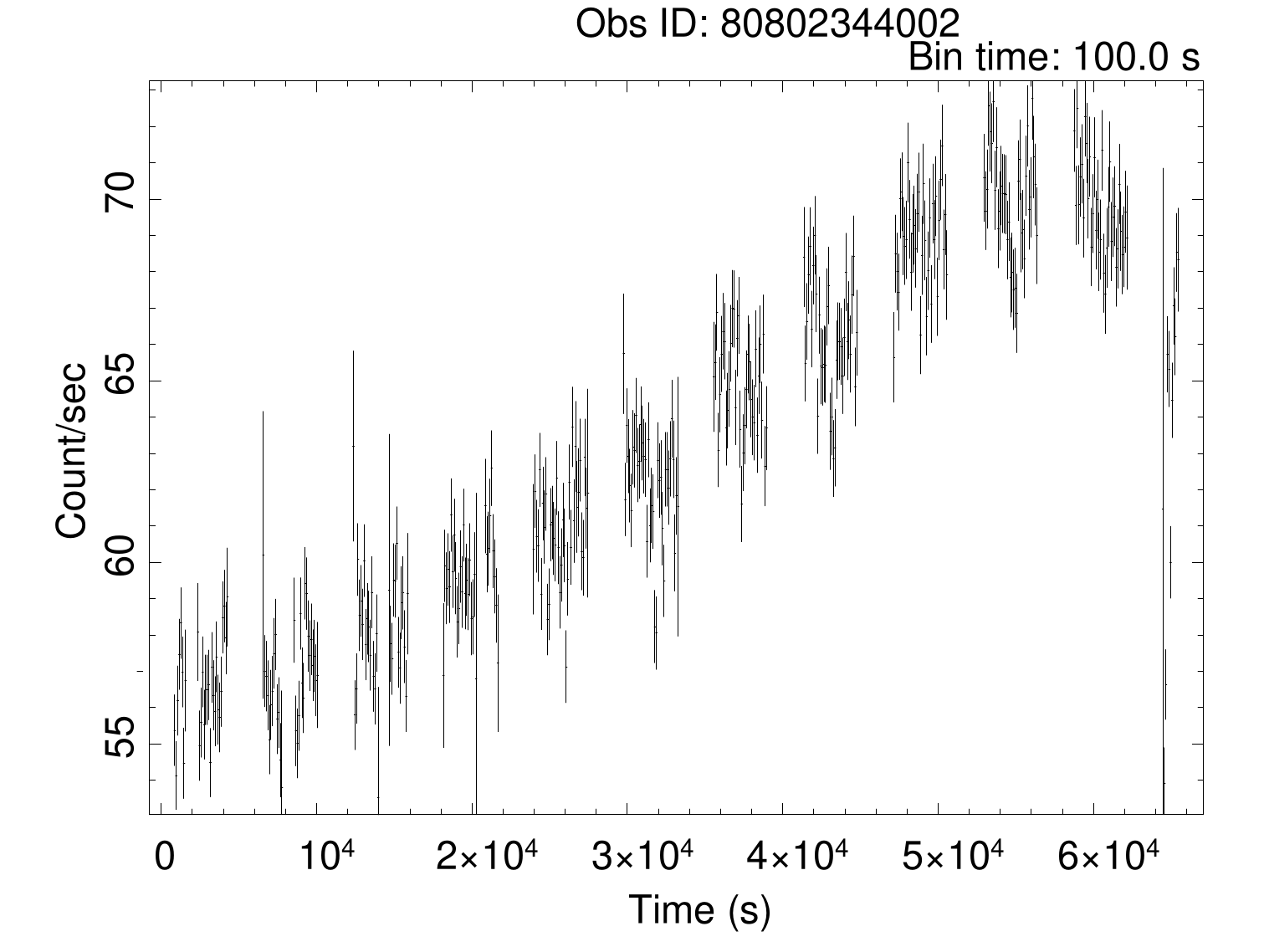} 
\includegraphics[height=0.42\textwidth, angle=-90]{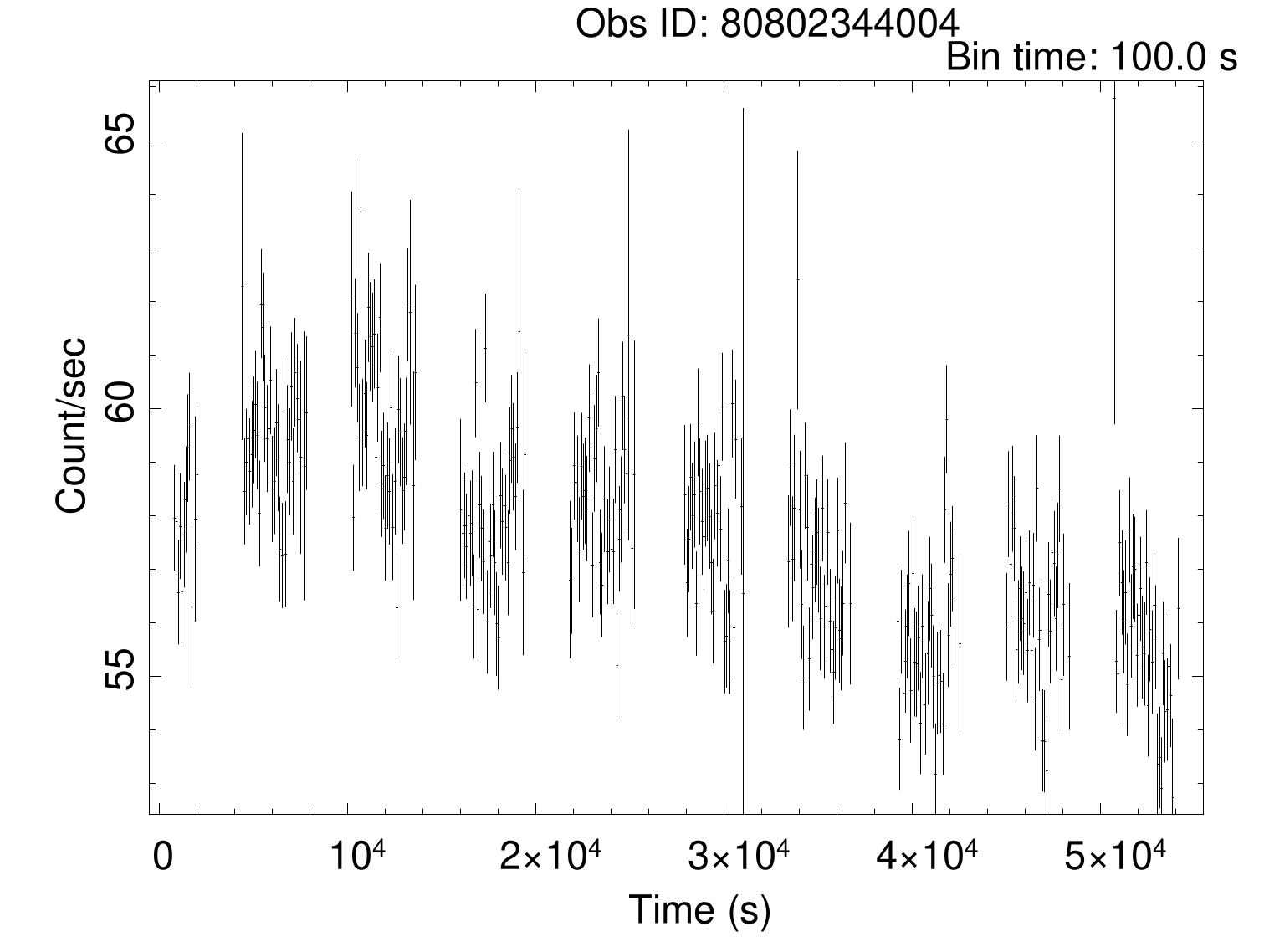}
\caption{\emph{NuSTAR} lightcurve (3.0-79.0 keV) for observations corresponding to MJD 59942 and 59962.}
\label{nustar_lc}
\end{figure}

\emph{NuSTAR} \citep{Harrison2013} is a space-based observatory that contains two detector units (FPMA and FPMB) to image astrophysical objects in hard X-ray. The working energy range of focusing telescopes is from 3 to 79 keV. We used \texttt{NuSTARDAS} pipeline \texttt{v0.4.9} and \texttt{CALDB v20220413} to reduce the \emph{NuSTAR} data. DS9 software was used to choose a circular source region of radius 120 arcsec at the source location and a circular region outside the source region with the same area for the background. Then, \texttt{NUPRODUCTS} tool was used to extract spectra and lightcurve files for both modules. Before spectral fitting, we rebin the source spectra for a minimum of 100 counts per bin using \texttt{GRPPHA} task. The \emph{NuSTAR} lightcurve for observation ID 80802344002 and 80802344004 are shown in figure \ref{nustar_lc}.

\begin{table}
\caption{Observation details of XTE J2012+381 analyzed in this work. The details of the source region chosen to extract lightcurve (count rate) are given in Section 2.}
\begin{tabular}{ccccc}
\hline
\begin{tabular}[l]{@{}l@{}}Time\\ (MJD) \end{tabular} & Instrument & Obs ID      & \begin{tabular}[c]{@{}c@{}}Exposure\\ (ks)\end{tabular} & count rate \\ \hline
59942 & NuSTAR     & 80802344002 & 29.1   & 64                                                 \\
           & Swift/XRT  & 00089581002 & 1.7   &  80                                              \\ \hline
59962 & NuSTAR     & 80802344004 & 23.2     & 58                                              \\
           & Swift/XRT  & 00015473007 & 1.3   & 49                                                 \\ \hline
\end{tabular}
\label{data}
\end{table}

\begin{table*}
\caption{Summary of all models used in Spectral fits. Note: M1.1 is an initial model used to take care of the disk, corona, and reflection components. M1.2 uses a high-density reflection model \texttt{reflionx} with disk density extending up to $10^{22}\;\mathrm{cm}^{-3}$, M1.3 is the Cp variant of the relxill model; it also provides density variation up to $10^{20}\;\mathrm{cm}^{-3}$. M1.3a and M1.3b correspond to fitting using fixed emissivity indices and varying emissivity indices, respectively. M1.4 is a lamppost variant of model M1.2; M1.5 is a lampost variant of model M1.3; M2.1 is an initial model to fit soft states; and M2.1 is a relativistic version of model M2.1. M2.2a and M2.2b corresponds to required parameter values taken from models M1.3b and M1.5, respectively. The model \texttt{constant} is a multiplicative model used to take care cross-calibration difference between Swift/XRT and NuSTAR FPMA/B.}
\begin{tabular}{cc}
\hline
\hline
Model & Components\\
\hline
M1.1 & \texttt{TBabs*(relxill + nthcomp + diskbb)}        \\
M1.2 & \texttt{TBabs*(nthcomp + diskbb + relconv*reflionx\_HD\_nthcomp\_v2.fits)}    \\
M1.3 & \texttt{TBabs*(relxillCp + nthcomp + diskbb)}         \\
M1.4 & \texttt{TBabs*(nthcomp + diskbb + relconv\_lp*reflionx\_HD\_nthcomp\_v2.fits)}  \\   
M1.5 & \texttt{TBabs*(relxilllpCp + nthcomp + diskbb)}   \\ 
M2.1 & \texttt{TBabs*(pow+diskbb)}     \\ 
M2.2 & \texttt{TBabs*(thcomp*kerrbb)}    
\\ \hline
\end{tabular}
\label{modelsummarytable}
\end{table*}

\section{Spectral analysis and Results}
The \emph{Swift/XRT} and \emph{NuSTAR} spectra in the energy range 1.0–78.0 keV were modelled using the \texttt{XSPEC v12.13.0c} software package \citep{Arnaud1996}. We used the initial spectral model, M1.1 : \texttt{constant* TBabs*(relxill + nthcomp + diskbb)} to fit the data during MJD 59942. 
The model \texttt{constant} is used to reconcile the calibration difference between \emph{Swift/XRT} and \emph{NuSTAR FPMA/B}. The model \texttt{TBabs} takes care of the absorption due to interstellar medium with cross-sections and abundances given in \citet{Wilms2000}. The equivalent hydrogen column $(n_H)$ is the parameter of the model. The \texttt{diskbb} spectral model describes the thermal emission from the accretion disk as multiple blackbody components \citep{Mitsuda1984,Makishima1986}. The temperature at inner disk radius $(T_{in})$ and norm ($=(R_{in}/D_{10})^2cos\theta$, where $R_{in}$ is ``an apparent” inner disk radius in km, $D_{10}$ the distance to the source in units of 10 kpc, and $\theta$ the angle of the disk) are the parameters of diskbb. 
The model \texttt{nthcomp} describes the thermal comptonization of seed photons at temperature $kT_{bb}$ by thermal electrons at temperature $kT_e$ \citep{Zdziarski1996,Zycki1999}. The seed photon distribution can be blackbody photons or thermal photons from the accretion disk. 
The \texttt{relxill} model \citep{Garcia2014,Dauser2013} computes the relativistic reflection spectrum from an illuminated accretion disk that surrounds a compact object. The \texttt{relxill} model is a combination of the ionized reflection model \texttt{xillver} \citep{Garcia2010,Garcia2011,Garcia2013} and the relativistic blurring kernel \texttt{relline} \citep{Dauser2010,Dauser2014}. The \texttt{relxill} model contains inner emissivity index $(q_1)$, outer emissivity index $(q_2)$, break frequency $R_{br}$, spin of black hole $(a)$, inclination angle $(i)$, inner radius of the accretion disk (Rin), the outer radius of the accretion disk (Rout), redshift to the source (z), Power law index $(\Gamma)$, log of ionization of accretion disk (log$\xi$), iron abundance (AFe), high energy cutoff of the primary spectrum (Ecut), reflection fraction parameter (refl\_frac) and norm. The log$\xi$ ranges from 0 to 4.7 which corresponds to neutral to highly ionized. The model \texttt{gaussian} was used to account for the Au-M edge at 2.35 keV due to instrumentation. 
The \texttt{relxill} model assumes a power-law continuum with an exponential cutoff at high energies. The upper energy limit of NuSTAR is 78 keV and no cutoff was observed upto this energy range in this data. So, in order to account for the power-law continuum, we employed the physical model \texttt{nthcomp} instead of exponential cutoff model. 

\begin{figure*}
\centering
\subfloat {{\includegraphics[height=0.38\textwidth, angle=0]{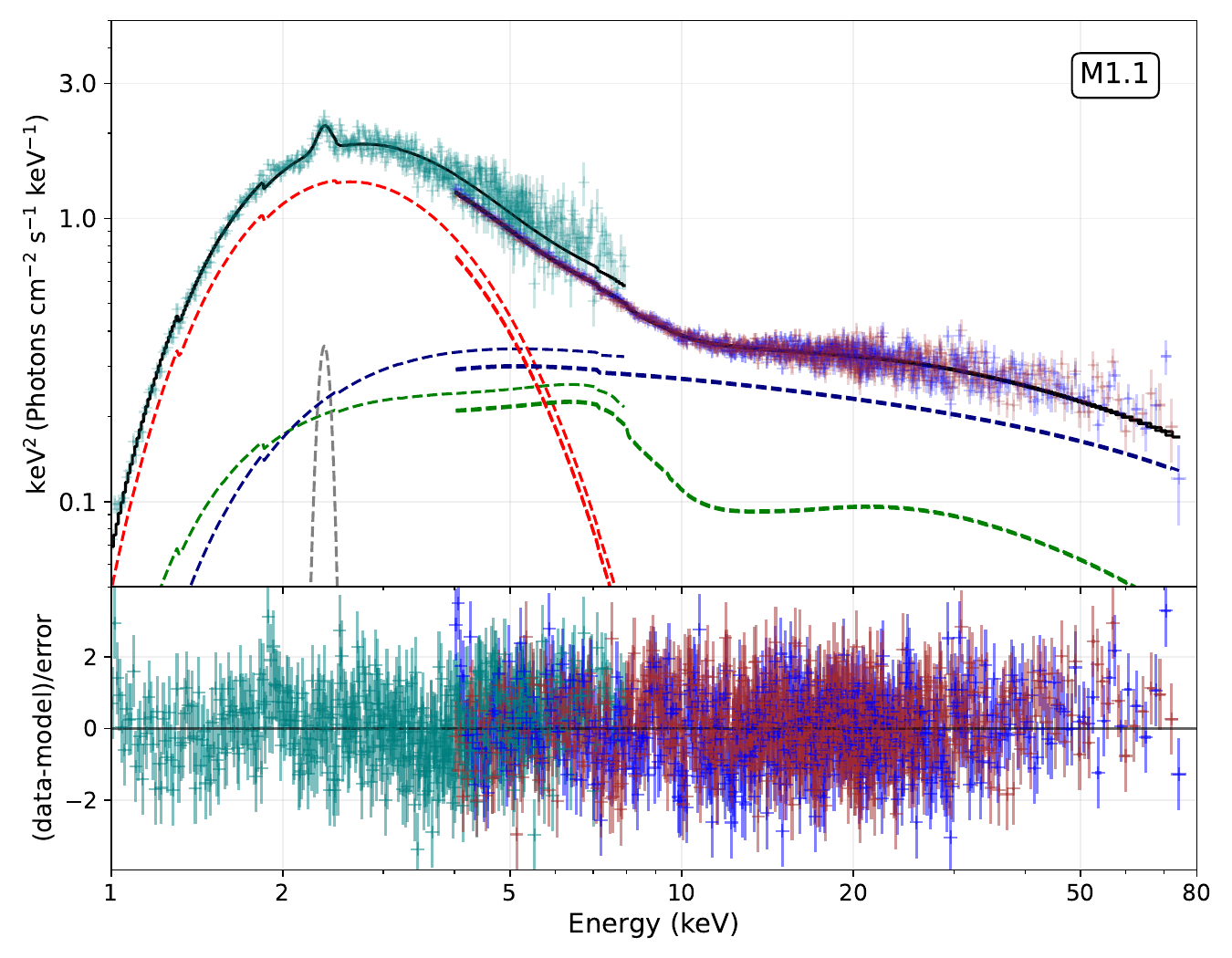} }}
\subfloat {{\includegraphics[height=0.38\textwidth, angle=0]{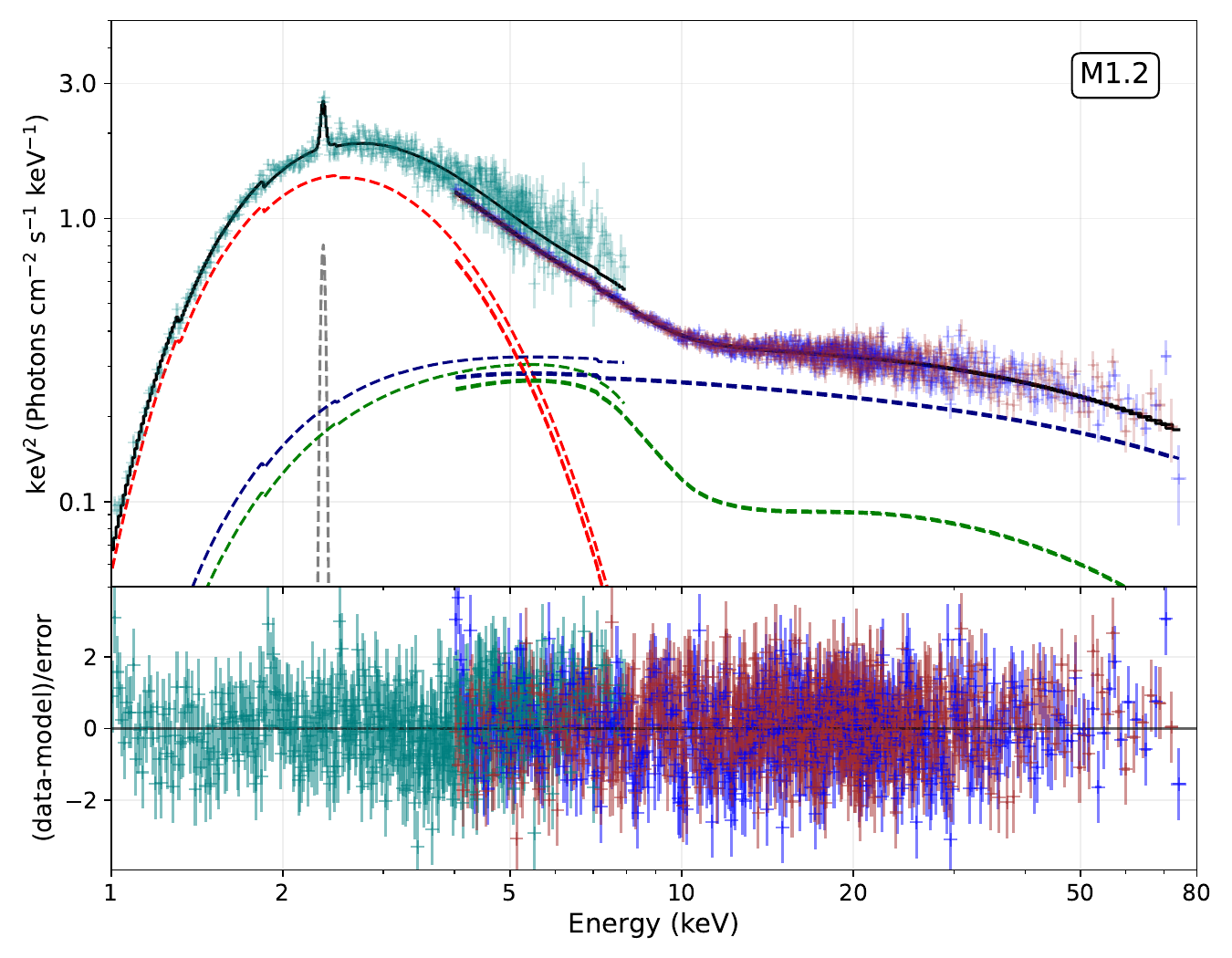} }}
\qquad
\subfloat {{\includegraphics[height=0.38\textwidth, angle=0]{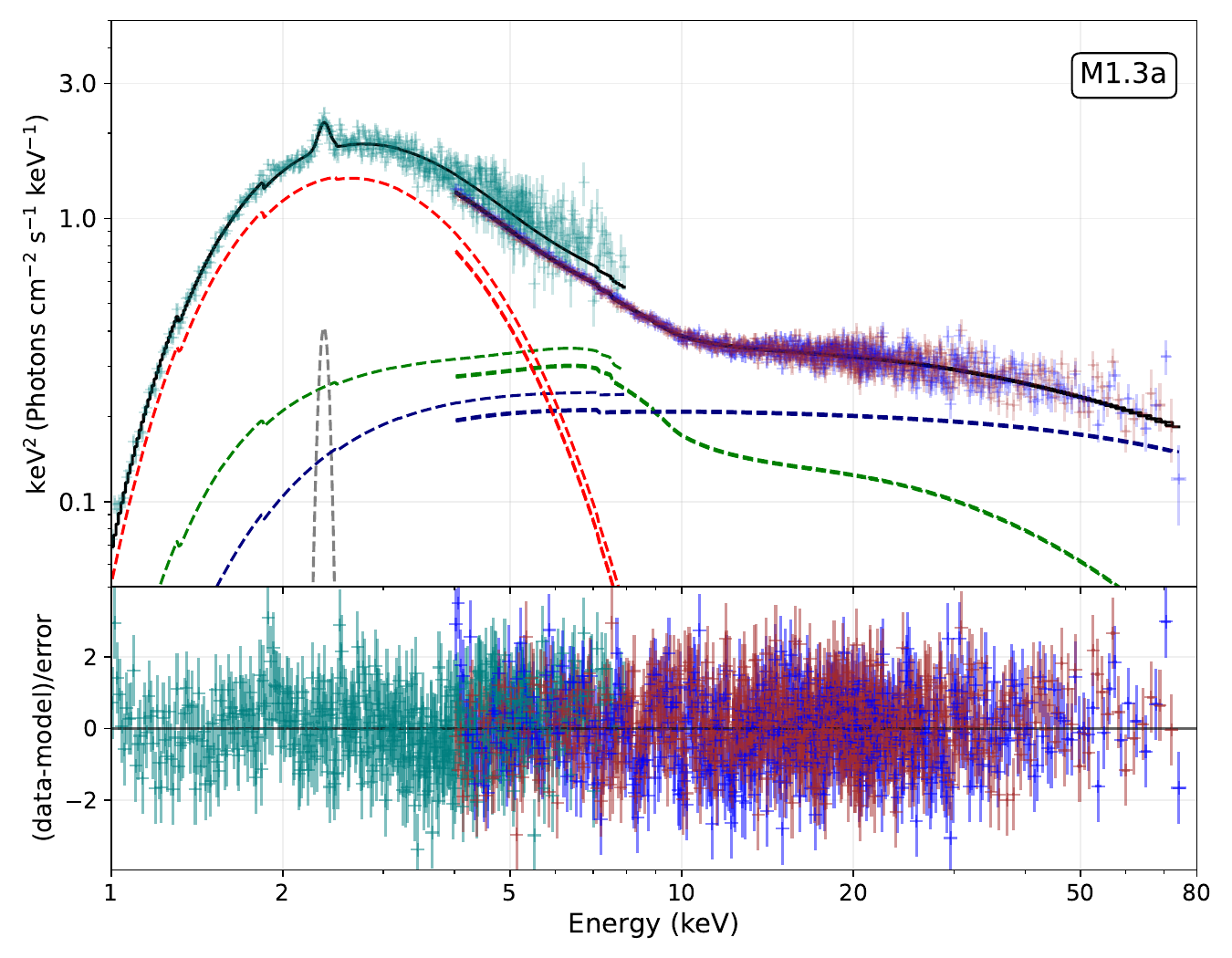} }}
\subfloat {{\includegraphics[height=0.38\textwidth, angle=0]{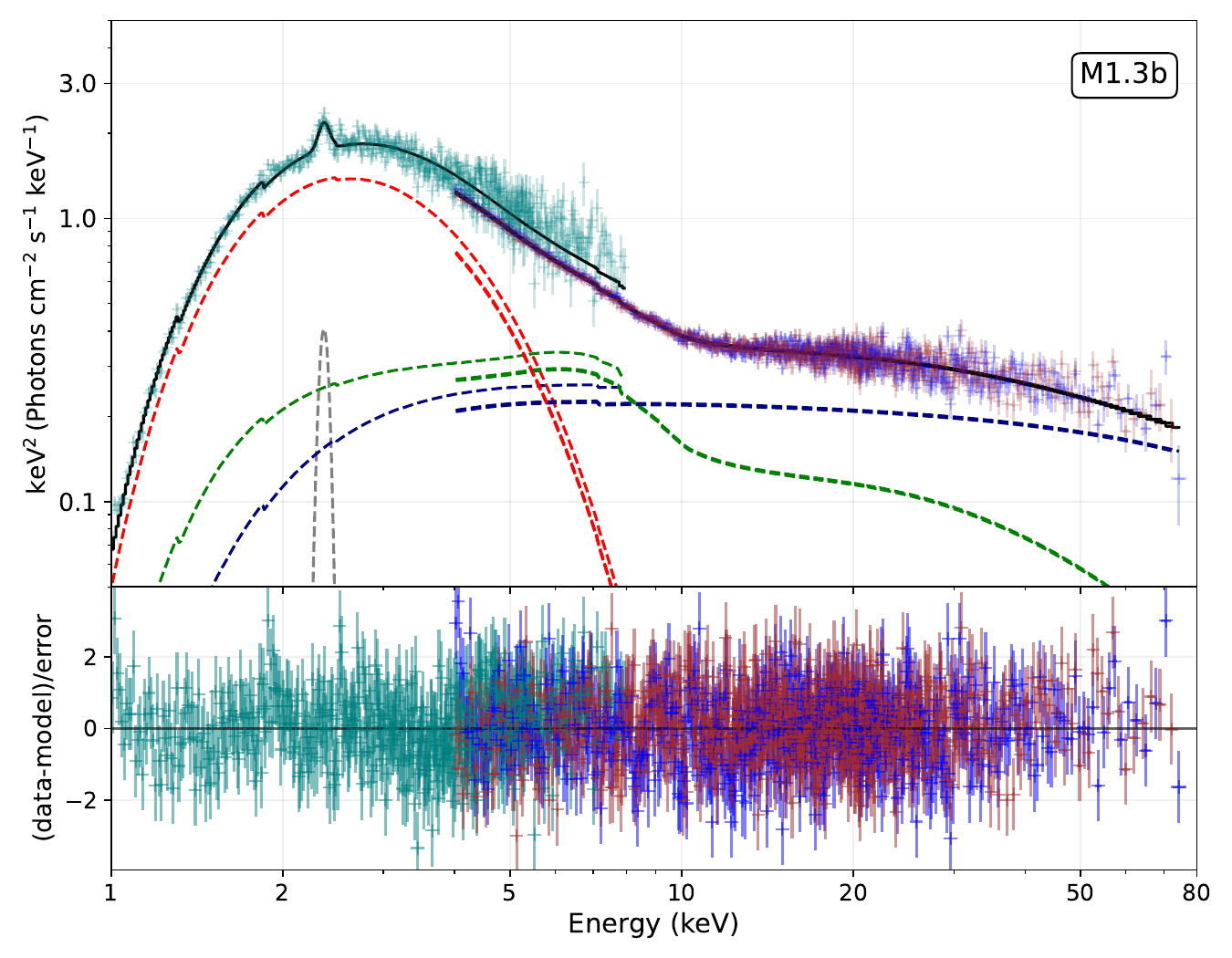} }}

\qquad
\subfloat {{\includegraphics[height=0.38\textwidth, angle=0]{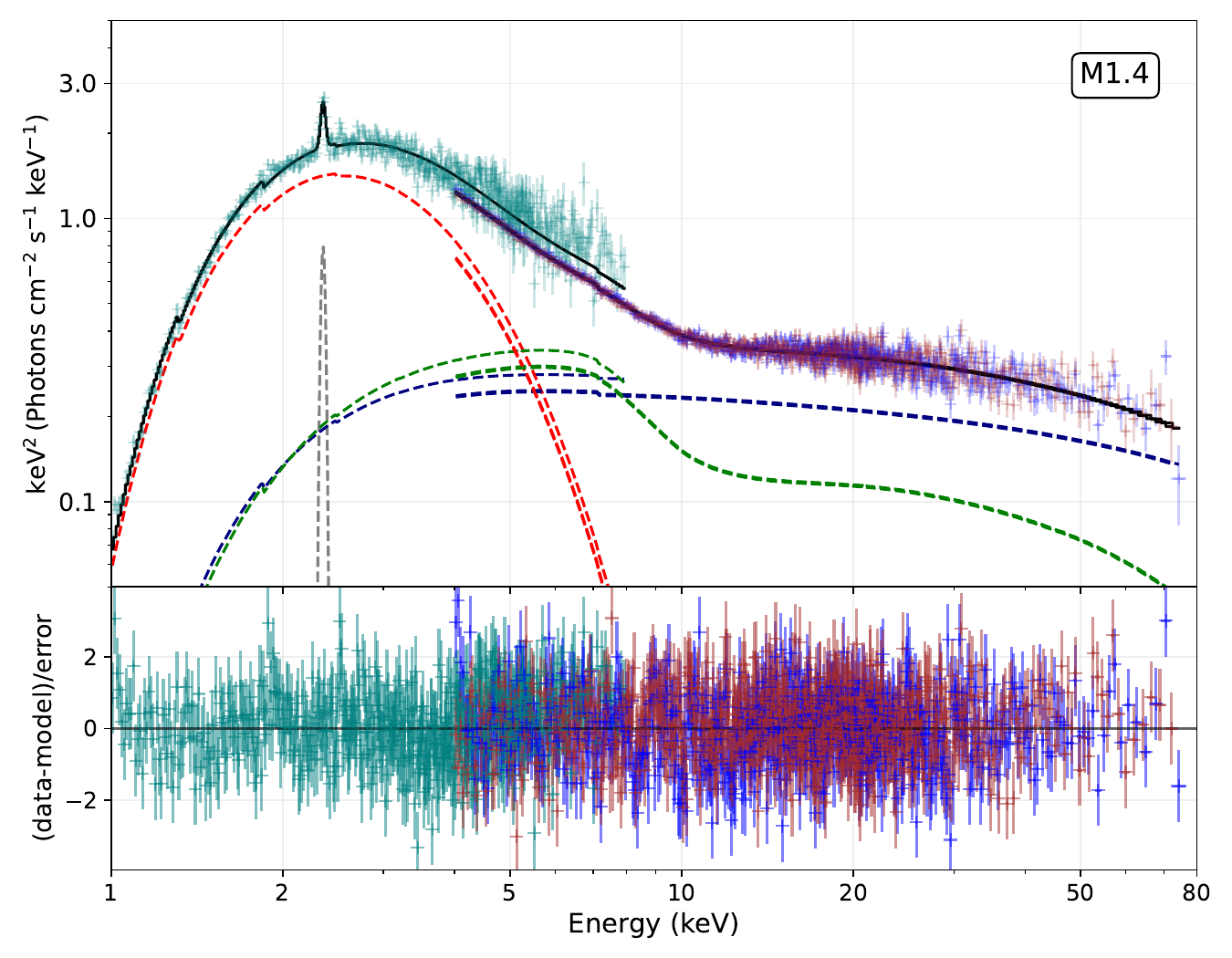} }}
\subfloat {{\includegraphics[height=0.38\textwidth, angle=0]{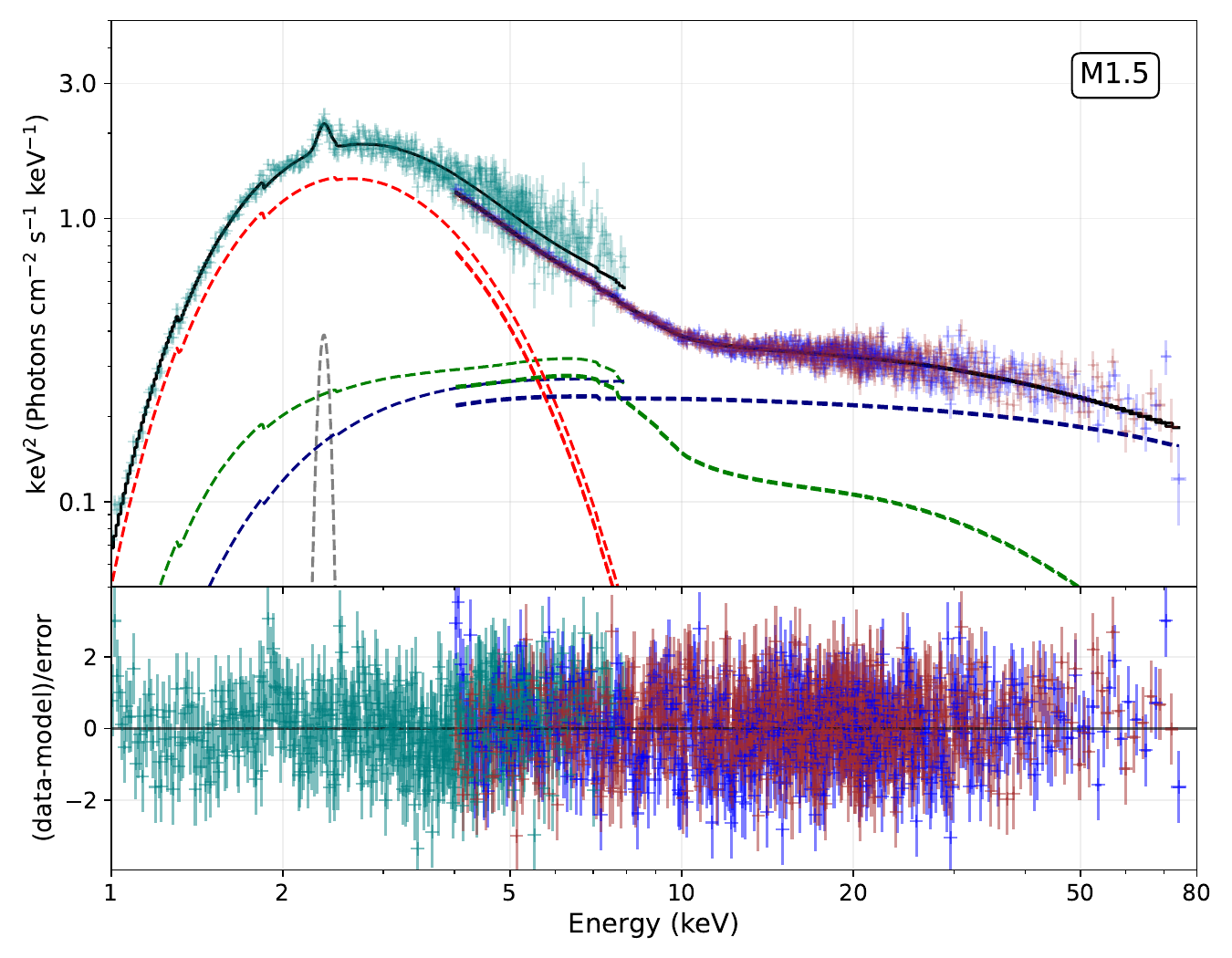} }}

\caption{Spectral fitting of \emph{Swift/XRT} and \emph{NuSTAR} data for MJD 59942. The details of models is given in table \ref{modelsummarytable} and their fitted  parameters value is given in table \ref{parameters} and table \ref{reflionx}. The disk component, corona component and reflection component are shown by red, blue and green dotted lines respectively. The energy range $1.0-8.0\;\mathrm{keV}$ and $4.0-78.0\;\mathrm{keV}$ was used to fit the \emph{Swift/XRT} and \emph{NuSTAR} data respectively.}
\label{spectrum42}
\end{figure*}

\begin{figure}
\centering
\subfloat {{\includegraphics[height=0.38\textwidth, angle=0]{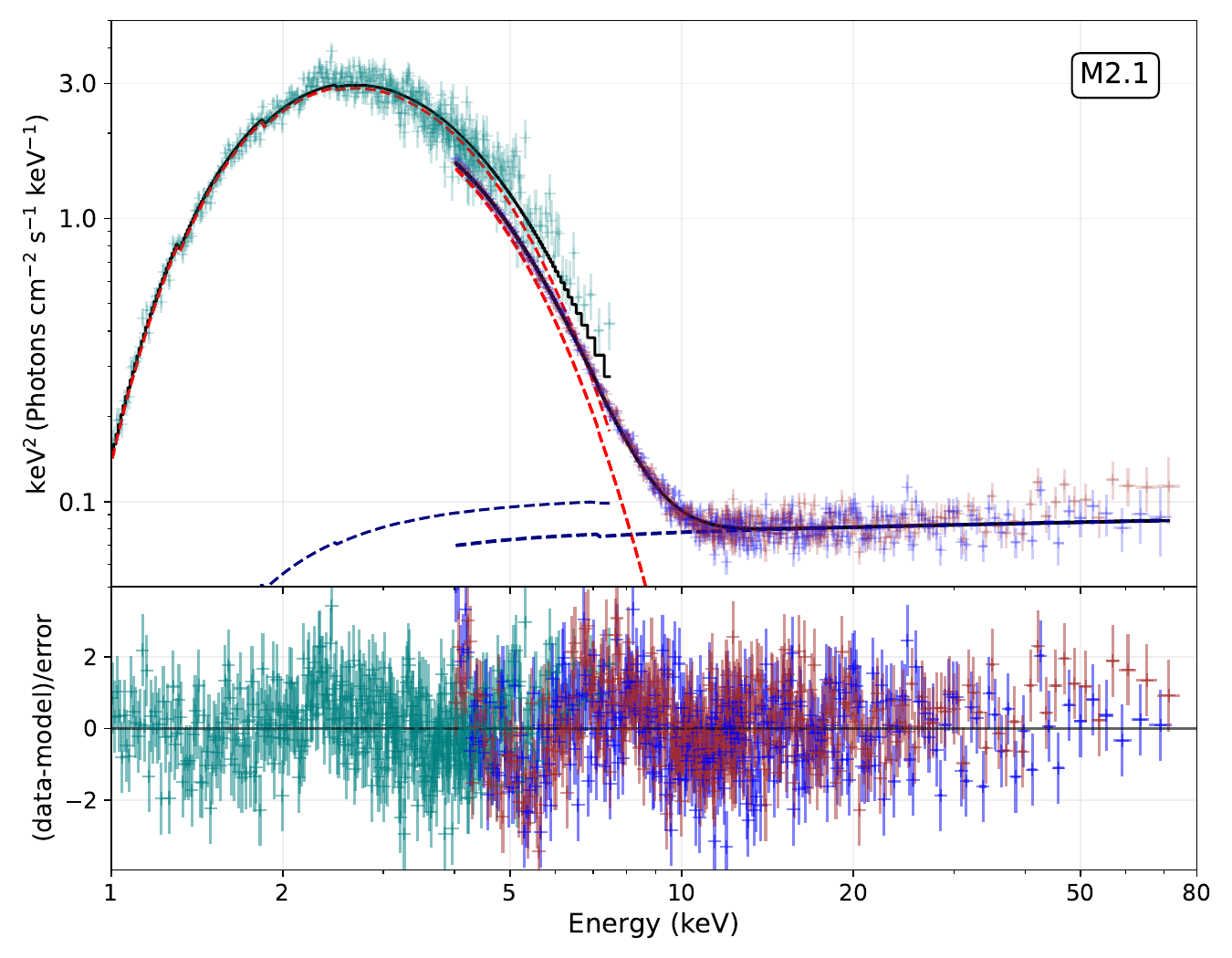} }}
\qquad
\subfloat {{\includegraphics[height=0.38\textwidth, angle=0]{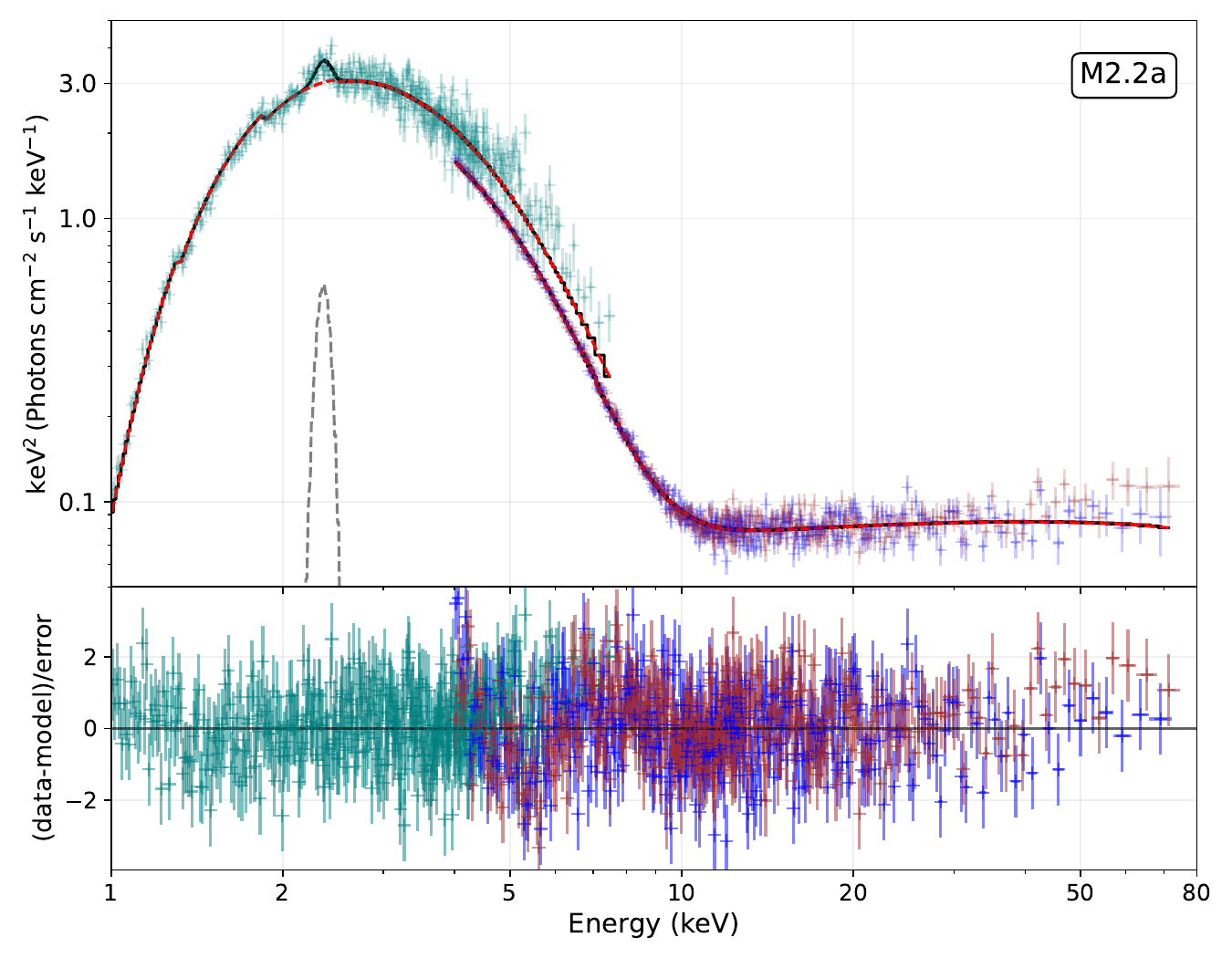} }}
\qquad
\subfloat {{\includegraphics[height=0.38\textwidth, angle=0]{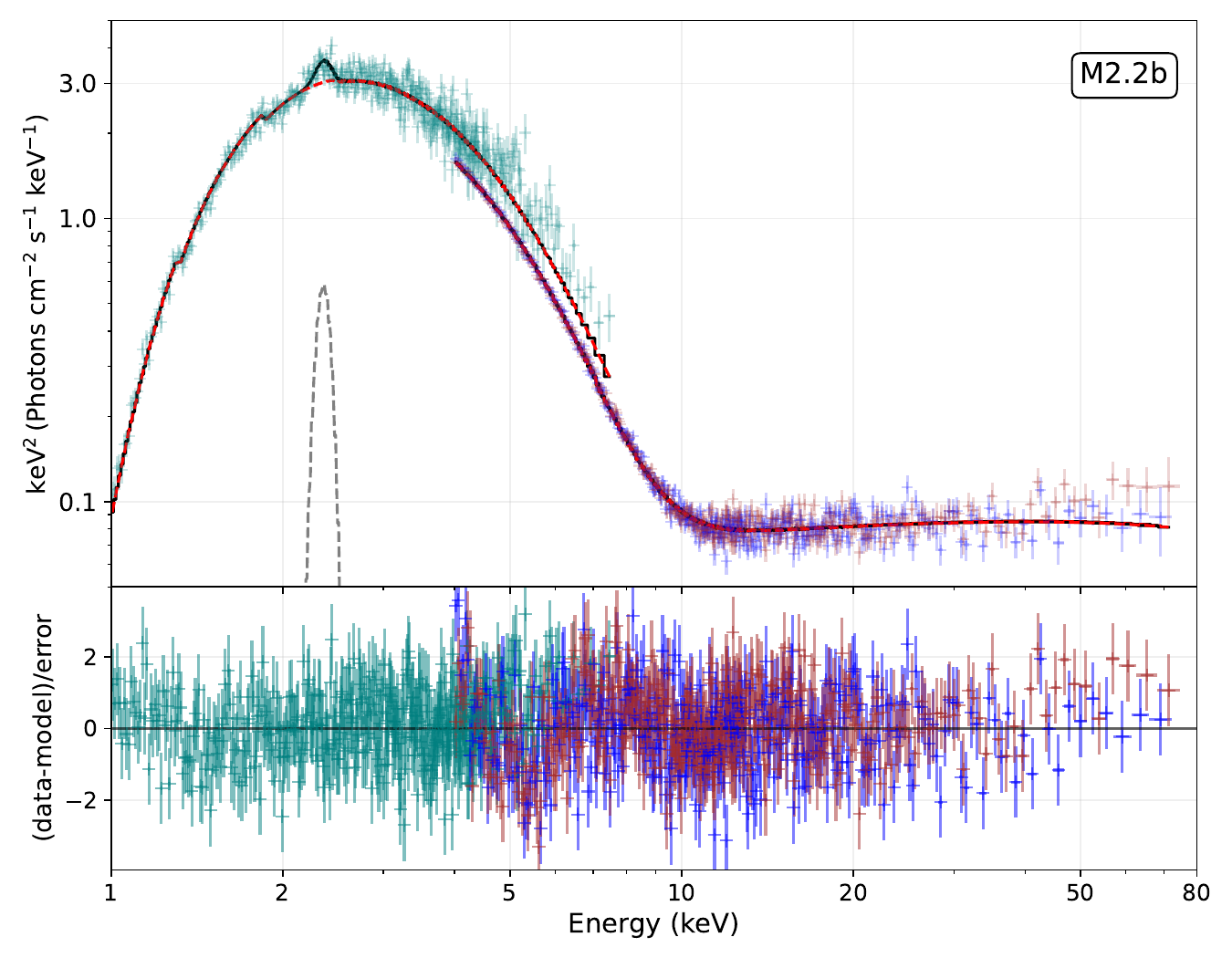} }}

\caption{Spectral fitting of \emph{Swift/XRT} and \emph{NuSTAR} data for MJD 59962. The details of models is given in table \ref{modelsummarytable} and their fitted parameters value is given in table \ref{mjd59962}. The energy range $1.0-8.0\;\mathrm{keV}$ and $4.0-78.0\;\mathrm{keV}$ was used to fit the \emph{Swift/XRT} and \emph{NuSTAR} data respectively.}
\label{spectrum62}
\end{figure}

To model the spectrum, the emissivity indices in \texttt{relxill} were kept fixed at the default value of 3.
The \texttt{relxill} model used to fit the reflection component only by setting refl\_frac = -1. We set the $R_{\mathrm{in}} = R_{\mathrm{ISCO}}$. The temperature of the thermal electrons was fixed at $100\;\mathrm{keV}$. The temperature of Ecut was fixed at 300 keV by using the relation $\mathrm{Ecut}=3kT_{e}$. We link the photon index ($\Gamma$) of \texttt{relxill} and \texttt{nthcomp}; and seed photon temperature ($kT_{bb}$) of \texttt{nthcomp} with inner disk temperature $(T_{in})$ of \texttt{diskbb}. The fitting of the data yielded a goodness-of-fit $\chi^2$/dof = 1763/1523. The fitting process gave the spin of the black hole, $a=0.932_{-0.076}^{+0.047}$ and the inclination of the source with respect to the normal to the accretion disk, $i=51.2^{\circ\;+1.3}_{-1.3}$. The disk temperature came out to be $T_{\mathrm{in}}=0.78\pm0.01\;\mathrm{keV}$. Among other parameters of the model, $n_{\mathrm{H}}$ , the equivalent neutral hydrogen column density came out to be $=(1.39\pm0.02)\times10^{22}\;\mathrm{cm^{-2}}$ and the asymptotic photon spectral index, $\Gamma=2.20\pm0.01$.
Our fitting result shows that the accretion disk is highly ionized with $\mathrm{log}\xi=4.34_{-0.09}^{+0.10}$. The details of the fitted parameters are tabulated in Table \ref{parameters}. It is worth noting that the fitting result shows a supersolar iron abundance with $A_{\mathrm{Fe}} > 8.39\:A_{\mathrm{Fe,\odot}}$.

The reflection model \texttt{relxill} consider the low disk density ($n_e \sim 10^{15}\;\mathrm{cm^{-3}}$). However, the standard $\alpha$-disc model \citep{SS1973} and 3D magneto-hydrodynamic (MHD) simulations \citep{Noble2010, Schnittman2013} predict larger disk density for stellar mass black holes. Some studies have shown that high disk density models may bring down the iron abundance. To address the issue of high iron abundance, We fitted the data with the model, M1.2 : \texttt{constant*TBabs*(nthcomp + diskbb + relconv*reflionx\_HD\_nthcomp\_v2.fits)}. The \texttt{reflionx\_HD\_nthcomp\_v2.fits}\footnote{https://www.michaelparker.space/reflionx-models} is a rest frame disk \texttt{reflion} model calculated with the reflionx code \citep{Ross2005} which assumes primary spectrum as \texttt{nthcomp}. It is a high-density reflection model with disk density extending upto $10^{22}\;\mathrm{cm^{-3}}$. Other free parameters of the reflection model include the ionization parameter $\xi=L/nr^2$ (Where $L$ is ionization luminosity from the primary source, $n$ is the density and $r$ is the distance from the ionizing source) and the iron abundance ($A_\mathrm{Fe}$). The \texttt{relconv} is a convolution kernel that required to include the relativistic effects \citep{Dauser2010}. This model contains the following parameters: black hole spin ($a$), the disk inclination ($i$), the inner disk radius ($R_{in}$) and the emissivity profile. The photon index ($\Gamma$) and coronal temperature parameters in \texttt{nthcomp} were tied to the same parameters used in the model \texttt{reflionx\_HD\_nthcomp\_v2.fits}. The spin value was fixed at 0.932 as measured by \texttt{relxill} model (M1.1). The $R_{in}$ fixed at $R_{ISCO}$. Newtonian emissivity profile ($q_1=q_2=3$) was taken to fit the data. The inclination angle was left free during the fit. The spectral fitting parameters are given in table \ref{reflionx}. The fit gave a high value of disk density $(n_e/\mathrm{cm}^{-3}) > 4.9\times10^{21}$. The iron abundance of disk came out to be $A_\mathrm{Fe}=2.51_{-1.46}^{+1.48}\:A_{\mathrm{Fe,\odot}}$. The ionization parameter of the disk came out to be $2068_{-710}^{+914}\;\mathrm{erg\; cm\;s^{-1}}$. The details of this fit are provided in Table \ref{reflionx}. Here we could not determine the value of the spin of the black hole, but we estimated its upper limit, $a\lesssim$~0.7.

The \texttt{reflionx} based reflection model confirmed the high disk density. So, we explored the high disk density model \texttt{relxillCp} (a \texttt{relxill} based reflection model) to estimate the spin of BH and the inclination of the source. We replaced \texttt{relxill} model with \texttt{relxillCp} model in M1.1. The model becomes, M1.3 : \texttt{constant* TBabs*(relxillCp + nthcomp + diskbb)}. The relxillCp model uses primary spectrum as \texttt{nthcomp} instead of \texttt{cutoffpl}. It allow to vary disk density in a range of $n = 10^{15}-10^{20}\;\mathrm{cm}^{-3} $. It has an additional parameter logN for disk density and uses corona temperature $kT_e$ instead of Ecut. Initially, we fixed the emissivity profile at Newtonian values and logN at 20. The value of spin and inclination came out to be $0.887_{-0.112}^{+0.034}$ and $37.9_{-2.0}^{+2.8}$ respectively (This model is tabulated as M1.3a in Table \ref{parameters}). The other parameters did not vary much. A reflection model that prefers high spin is expected to have a relatively steep emissivity profile \citep{Wilkins2012}. So, we free the inner emissivity index and fixed the outer emissivity index at 3 (This model is tabulated as M1.3b in Table \ref{parameters}). The $q_1$ is came out to be $7.4_{-2.34}^{P}$ and $R_{br}$ came out to be $3.95_{-0.59}^{+1.93}\;R_g$. The upper bound of $q_1$ was pegged at 10. The fitting process gave the spin of BH and inclination of source $0.883_{-0.061}^{+0.033}$ and $46.2_{-2.0}^{+3.7}$ degrees, respectively. In both cases, The fitting resulted in super solar iron abundance.

Since the geometry of disc-corona is not known, we explored the lamp-post geometry model to assess the effect on inclination and spin. We first checked the disk density using the lamp post geometry model. The \texttt{relconv} model was replaced by the \texttt{relconv\_lp} model in M1.2. The model becomes, M1.4: \texttt{constant*TBabs*(nthcomp + diskbb + relconv\_lp*reflionx\_HD\_nthcomp\_v2.fits)}. The \texttt{relconv\_lp} is model with lamppost geometry. It replaces the emissivity profile parameter with the height of the coronal source. Compared to M1.1, M1.3b offered a far better fit. Thus, we fixed the black hole's spin at 0.883 in this fit.  The fitting gave the disk density $n_e>3.11\times10^{21}\;\mathrm{cm^{-3}}$ and iron abundance came out to be $2.12_{-0.50}^{+1.01}\:A_{\mathrm{Fe,\odot}}$.

Finally, we replaced the \texttt{relxill} model with \texttt{relxilllpCp} model in M1.1. The model becomes, M1.5: \texttt{constant* TBabs*(relxilllpCp + nthcomp + diskbb)}. The \texttt{relxilllpCp} is a most powerful model from \texttt{relxill} family. It predicts the emissivity from the lamp post geometry and primary spectrum modelled by \texttt{nthcomp}. It contains the following parameters: inclination angle $(i)$, spin of black hole $(a)$, inner radius of the accretion disk (Rin), Outer radius of the accretion disk (Rout), the height of the primary source above the black hole (h), the velocity of primary source (beta), Power law index of primary source spectrum $(\Gamma)$, ionization of the accretion disk $(log\xi)$, logarithmic value of the density (logN), Iron Abundance (Afe), electron temperature in the corona (kTe), reflection fraction parameter (refl\_frac), redshift to the source (z), ionization gradient index (iongrad\_index), parameter to specify the radial behaviour of the ionization on the accretion disk (ion\_grad\_type), switch\_reflfrac\_boost, switch\_returnrad and Norm. This model allows for varying ionization and density gradients in the disk. It also allows to vary logN from 15 to 20. 

For simplicity, we took the stationary primary source (beta=0), and constant ionization and constant density (grad\_type=0). The grad\_index parameter is set at zero. We fixed logN at 20 due to high disk density. Assuming Rin=$R_{ISCO}$, the Rin was fixed at -1. The Rout was fixed at a default value of 400 Rg (Rg = GM/$c^2$). We switch the returning radiation on (switch\_returnrad=0) and set the switch\_reflfrac\_boost=1 (refl\_frac behave as the reflection fraction). The refl\_frac set =-1 to return the reflection only from \texttt{relxilllpCp} model. The kTe and $\Gamma$ of \texttt{nthcomp} and \texttt{relxillCp} were linked. The kTe was fixed at a reasonable value of 100 keV. All other parameters were allowed to vary freely including the height of the corona above the black hole (h), inclination angle ($i$), spin of BH ($a$), ionization of the accretion disk (log$\xi$), iron abundance (Afe), and normalization of \texttt{relxilllpCp}. The model parameters are listed in detail in Table \ref{parameters}. The spectral fittings for this observation are shown in Figure \ref{spectrum42}. The best fit gave the height of the corona of $2.15_{-0.28}^{+0.58}\;R_g$. 
This gives the spin of the black hole and inclination of source $0.892_{-0.044}^{+0.020}$ and $43.1_{-1.2}^{+1.4}$ degrees, respectively.

To fit the spectra for MJD 59962, we started with the model, M2.1: \texttt{constant*TBabs*(pow+diskbb)}. This resulted in the temperature of the accretion disk $T_{in}\sim0.829_{-0.002}^{+0.002}\;\mathrm{keV}$ and norm of \texttt{diskbb} $1002_{-18}^{+19}$. The photon index $\Gamma$ came out to be $1.96\pm0.02$. The $\chi^2/dof$ for the fit came out to be 1377/1113. Then we replaced the \texttt{diskbb} model by \texttt{kerrb} model and \texttt{power-law} by \texttt{thcomp} model. The \texttt{kerrbb} is a relativistic model that is used to take care of the disk spectrum \citep{Li2005} and \texttt{thcomp} is a convolution model for the thermal comptonization. So, the model is M2.2:\texttt{constant*TBabs*(thcomp*kerrbb)}. In this case, we used the spin and the inclination of the source as obtained by the fitting of the spectrum during MJD 59942. We applied \texttt{gain fit} command in \texttt{xspec} and additionally used a Gaussian Au-M edge at 2.35 keV. We extended the energy range (0.01-1000 keV) required by the response for \texttt{thcomp} model to work properly. \citet{Shimura1995} suggested that the value of $f_{\mathrm{col}}$ can span 1.5-1.9 for accretion disks around a stellar-mass black hole. A few studies provided a theoretical range of 1.4–2.0 for $f_{\mathrm{col}}$ in the high/soft state of X-ray binaries \citep{Davis2005,Davis2019}. We used the default value of $f_{\mathrm{col}}$ that is 1.7. The fitted parameters are given in Table \ref{mjd59962}. The model is tabulated as M2.2a for spin and inclination estimated from the \texttt{relxillCp} model and M2.2b for spin and inclination estimated from the \texttt{relxilllpCp} model. The spectral fittings for MJD 59962 using above mentioned models are shown in Figure \ref{spectrum62}. The summary of all spectral models used in this work is given in table \ref{modelsummarytable}.

\begin{table*}
\caption{Fitted parameters for the MJD 59942 for the models described in the text. * represents the parameter that is fixed. Here, $n_\mathrm{H}$: equivalent neutral hydrogen column density, $a$: the spin of the black hole, $i$: the angle between the axis of the disk and the line of sight, $T_{\mathrm{in}}$: disk temperature, $N_{\mathrm{dbb}}$: normalization of diskbb model, $\Gamma$: asymptotic power-law index of the Comptonized photon distribution,  $q_1,\;q_2$: emissivity index for the inner and outer region, $R_{\mathrm{br}}$: break radius where emissivity profile changes (measured in gravitational radii), h: height of corona in lamp post geometry model (measured in gravitational radii), N: electron disk density measured in $\mathrm{cm}^{-3}$, $\xi$: ionization of the accretion disc in units of $\mathrm{erg\;cm\;s^{-1}}$, $A_{\mathrm{Fe}}$: iron abundance of the material in the accretion disk. The confidence ranges for the error bars are 90\% CI.}
\begin{tabular}{ll|ccc|c}
\hline
Component       & Parameter & M1.1 & M1.3a & M1.3b & M1.5 \\ \hline
TBabs           & $n_\mathrm{H}\;(\times10^{22}\;\mathrm{cm^{-2}})$      &  $1.39_{-0.02}^{+0.03}$  &   $1.37_{-0.02}^{+0.02}$  &   $1.39_{-0.01}^{+0.02}$  &  $1.38_{-0.01}^{+0.01}$  \\
Diskbb          & $T_\mathrm{in}\;(\mathrm{keV})$      &  $0.78_{-0.01}^{+0.01}$  & $0.78_{-0.01}^{+0.01}$    &  $0.78_{-0.01}^{+0.01}$   &  $0.78_{-0.01}^{+0.01}$  \\
                & $N_{dbb}$      &  $756_{-53}^{+42}$  & $735_{-47}^{+44}$    &  $779_{-53}^{+51}$   & $743_{-44}^{+45}$   \\
NTHCOMP         & $\Gamma$     &  $2.20_{-0.01}^{+0.01}$  & $2.04_{-0.03}^{+0.04}$ &  $2.07_{-0.01}^{+0.01}$   &  $2.06_{-0.01}^{+0.01}$  \\
                & norm      & $0.16_{-0.02}^{+0.01}$   &  $0.10_{-0.02}^{+0.02}$   &  $0.11_{-0.03}^{+0.01}$   &  $0.11_{-0.01}^{+0.01}$  \\
relxill/Cp/lpCp & $q_1$        & 3*   &  3*   & $7.4_{-2.3}^{+P}$  &  -  \\
                & $q_2$        & 3*  &   3*  &  3*   &  - \\
                & $R_{br}\;(R_g)$       & 15*   &  15*   & $3.95_{-0.59}^{+1.93}$     &  -  \\
                & $h\;(R_g)$  &  -  &  -   &   -  &  $2.15_{-0.28}^{+0.58}$  \\
                & $a$         & $0.932_{-0.077}^{+0.047}$ & $0.887_{-0.122}^{+0.034}$ & $0.883_{-0.061}^{+0.033}$ & $0.892_{-0.044}^{+0.020}$  \\
                & $i \mathrm{(deg)}$        &  $51.2_{-1.3}^{+1.3}$   &  $37.9_{-2.0}^{+2.8}$& $46.2_{-2.0}^{+3.7}$   &  $43.1_{-1.2}^{+1.4}$ \\
                & Log N     & 15*   & 20*    &  20*   & 20*   \\
                & $\mathrm{log} \xi$        &  $4.34_{-0.09}^{+0.10}$   &  $3.86_{-0.17}^{+0.11}$ &$3.70_{-0.01}^{+0.15}$   &  $3.73_{-0.05}^{+0.13}$   \\
                & $A_{Fe}\;\mathrm{(solar)}$       &  $>8.39$  &   $>8.40$  &  $>8.97$   &   $>8.80$ \\
                & Norm $(\times10^{-3})$     &  $3.80_{-0.47}^{+0.54}$  &  $1.83_{-0.29}^{+0.30}$   &  $2.12_{-0.26}^{+0.49}$   & $40.4_{-11.53}^{+101.90}$   \\ \hline
                & $\chi^2$/dof  &  1764/1723   &  1741/1723   &  1726/1721   &  1736/1722  \\ \hline
\end{tabular}
\label{parameters}
\end{table*}

\begin{table}
\caption{Spectral Fitting parameters of the reflionx based reflection model used to estimate the disk density for the MJD 59942. }
\begin{tabular}{llcc}
\hline
Component           & Parameter & M1.2 & M1.4 \\ \hline
TBabs               & $n_\mathrm{H}\;(\times10^{22}\;\mathrm{cm^{-2}})$      & $1.37_{-0.02}^{+0.02}$ & $1.37_{-0.02}^{+0.02}$ \\
Diskbb              & $T_\mathrm{in}\;(\mathrm{keV})$      &  $0.75_{-0.01}^{+0.01}$  &  $0.75_{-0.01}^{+0.01}$  \\
                    & $N_{dbb}$      & $958_{-75}^{+89}$   &   $957_{-83}^{+79}$ \\
nthcomp             & $\Gamma$     & $2.15_{-0.06}^{+0.02}$   & $2.12_{-0.06}^{+0.04}$   \\
                    & norm      &  $0.15_{-0.05}^{+0.02}$  &  $0.13_{-0.04}^{+0.03}$  \\
Relconv(\_lp) & $q_1$        &   3* &  -  \\
                    & $q_2$        &   3* &  -   \\
                    & $R_{br}\;(R_g)$       &   15* & -   \\
                    & $h\;(R_g)$         &  -  &  $13.2_{-7.3}^{+17.4}$  \\
                    & $a$         & 0.932*   & 0.883*   \\
                    & $i$         &  $48.0_{-3.4}^{+1.4}$  & $39.5_{-8.4}^{+6.6}$   \\
Reflionx\_HD        & Density$\;(\mathrm{cm^{-3}})$  &  $>4.39\times10^{21}$  &  $>3.11\times10^{21}$  \\
                    & $\xi\;\mathrm{(erg\;cm\;s^{-1})}$        &  $2068_{-710}^{+914}$  &  $2240_{-761}^{+664}$  \\
                    & $A_{Fe}\;\mathrm{(solar)}$       & $2.51_{-1.46}^{+1.48}$   &  $2.12_{-0.50}^{+1.01}$  \\
                    & Norm      &  $0.293_{-0.007}^{+0.140}$  & $0.297_{-0.070}^{+0.202}$   \\ \hline
                    & $\chi^2$/dof  &  1732/1723  &  1728/1722\\ \hline
\end{tabular}
\label{reflionx}
\end{table}

\begin{table}
\caption{Fitted parameters for the MJD 59962 for the models described in the text. Here, $M_{BH}$: mass of the black hole and $\dot{M}$: mass accretion rate in units of $10^{18}\;\mathrm{g/s}$, $D$: distance of the source from the observer, cf: is covering fraction. The uncertainties in mass are reported as estimated in the text.}
\begin{tabular}{llccc}
\hline
Component & Parameter & M2.1 & M2.2a & M2.2b \\ \hline
TBabs     & $n_\mathrm{H}$     &  $1.22\pm0.02$    &   $1.56\pm0.06$    & $1.56\pm0.02$       \\
Diskbb    & $T_{in}\;(\mathrm{keV})$      & $0.83\pm0.01$     &  -     & -      \\
          & $N_{dbb}$      &  $1002_{-18}^{+19}$    &    -   & -      \\
Kerrbb    & $a$         &   -   &  0.883*  &  0.882*     \\
          & $i\;(\mathrm{deg})$         &   -   &    46.2*   &   43.1*    \\
          & $M_\mathrm{BH}\;(M_\odot)$       &  -    &  $7.95_{-3.25}^{+7.65}$     &  $7.48_{-2.75}^{+5.80}$     \\
          & $\dot{M}$      &  -    &   $0.31\pm0.01$    &  $0.30\pm0.01$      \\
          & $D_\mathrm{BH}\;(\mathrm{kpc})$       &  -    &   5.4*    &   5.4*    \\
Power law & $\Gamma$  &  $1.96\pm0.02$ &  -   &   -    \\
          & Norm      &  $0.07\pm0.01$    &   -   &   -    \\
thcomp    & $\Gamma$     &   -   &  $1.90\pm0.02$     &   $1.89\pm0.02$    \\
          & $kTe\;(\mathrm{keV})$       &    -  &   100*    &   100*    \\
          & $\mathrm{Cf}(\times10^{-2})$ &  -    &   $1.31\pm0.04$  &  $1.31\pm0.04$     \\ \hline
          & $\chi^2$/dof  &  1377/1113    &  1273/1113     &  1267/1113     \\ \hline
\end{tabular}
\label{mjd59962}
\end{table}

\section{Discussion and Conclusions}
In this work, we have analyzed the broadband X-ray spectrum (1.0-78.0 keV) of black hole X-ray binary XTE J2012+381 using simultaneous archival data from \emph{swift/XRT} and \emph{NuSTAR}. We fitted the continuum by \texttt{diskbb} or \texttt{kerrbb} model with the thermal comptonization model \texttt{nthcomp} or \texttt{thcomp}. For the reflection component, we employed the \texttt{relxill} based relativistic reflection models (\texttt{relxill}, \texttt{relxillCp}, and \texttt{relxilllpCp}) and \texttt{reflionx} based model (\texttt{reflionx\_HD\_nthcomp\_v2.fits}). The broadband X-ray spectrum fitting using observations on MJD 59942 was used to constrain the inclination of the disk and spin of the black hole. Furthermore, spectral fitting of observation on MJD 59962 was used to estimate the black hole mass.

 Accurate estimation of the ISCO size is crucial for black spin measurements using X-ray reflection spectroscopy \citep{Reynolds2014}.
In the intermediate state, considering the fact that the accretion disc extends upto ISCO, the spin of BH and inclination of the source were estimated using the relativistic reflection method. With the \texttt{reflionx} based reflection model, we found high disk density for XTE 2012+381. So, the data was fitted by taking logN=20 for \texttt{relxill} based reflection model \texttt{relxillCp}.
It is not true disk density as estimated by \texttt{reflionx} model M1.2 and M1.4. Since $R_\mathrm{in}$ is primarily constrained by the red wing of the broad iron line due to the gravitational redshift effect, which is independent of disk density. Therefore, using a lower value of disk density won't significantly change the spin estimate of BH.
By general relativistic ray-tracing simulations, \citet{Wilkins2012} showed that in the innermost parts of the disk, the emissivity profile falls off steeply, with the power-law indices ($q_1$) as high as $6\sim7$. \citet{Fabian2012} argued that the emissivity profile of the reflected emission from the innermost part of the disc must be fitted by steep power-law indices for a highly spinning black hole. In M1.3b, The inner emissivity profile is left to vary. It gave a steep inner emissivity profile with $q_1\sim7.4$. The BH spin and inclination were estimated to be $0.883_{-0.061}^{+0.033}$ and $46.2_{-2.0}^{+3.7}$ degrees, respectively.
The lamp post geometry model \texttt{relxilllpCp} gave the BH spin and inclination $0.892^{+0.044}_{-0.020}$ and $43.1_{-1.2}^{+1.4}$ degrees, respectively. Overall, this value of spin provided support for the validity of a highly spinning black hole in XTE J2012+381. 
\citet{Draghis2023} estimated the BH spin $a=0.988_{-0.030}^{+0.008}$ and inclination of accretion disk $i=68_{-11}^{+6}$ degrees by analysing the NICER and NuSTAR observations. These estimations were carried out utilizing a pipeline established by \citet{Draghis2023a}. In our study, We yielded a lower value of spin and inclination compared to that reported by \citet{Draghis2023}. One of the possible reason for this difference is that their final results depend on the \texttt{relxill} model only but we employed the Cp and lpCp variant models of \texttt{relxill} in our study.

The value of BH spin and inclination as estimated from \texttt{relxillCp} and \texttt{relxilllpCp} were used to estimate the BH mass using the ``continuum fitting" method by fitting the spectrum in the soft state. The BH mass estimation strongly depends on the value of spin, inclination, and distance of source from the observer. To estimate the uncertainty in BH mass, we estimated the mass at the extreme values of each parameter and calculated the mass range. By varying the inclination from 46.2 to 49.9 degrees, spin from 0.822 to 0.916 as estimated by \texttt{relxillCp} model and distance from 3.9 kpc to 8.79 kpc, we found the mass in the range 4.7 to 15.6 solar mass (The best-fit values are tabulated as M2.2a in Table \ref{mjd59962}). By varying the inclination from 41.9 to 44.5 degrees, spin from 0.848 to 0.912 as estimated by \texttt{relxilllpCp} model and distance from 3.9 kpc to 8.79 kpc, we found the mass in the range 4.73 to 13.28 solar mass (The best-fit values are tabulated as M2.2b in Table \ref{mjd59962}). The large uncertainty in mass is due to uncertainty in the distance. A more accurate measurement of the distance of the source in future will help to constrain the BH mass more accurately.

The measured iron abundance during MJD 59942 is found to be supersolar by \texttt{relxill} model. Supersolar iron abundance was observed in different black hole X-ray binaries, such as GX 339-4 \citep{Garcia2015, Parker2016}, V404 Cyg \citep{Walton2017}, Cyg X-1 \citep{Parker2015, Walton2016}, 4U 1543-47 \citep{Dong2020_1543, Prabhakar2023}, MAXI J1836-194 \citep{Dong2020_1836}, MAXI J1348-630 \citep{Chakraborty2021} and MAXI J0637-430 \citep{Jia2023}. \citet{Garcia2018} discussed the problem of the high iron abundance in accretion disks around black holes. Recently, \citet{Liu2023} analyzed the 21 spectra of six black hole X-ray binaries (MAXI J1535-571, GRS 1739-278, GS 1354-64, IGR J17091-3624, H 1743-322 and V404 Cyg) in hard state and found that 76\% of their sample require disk density higher than $10^{15}\;\mathrm{cm}^{-3}$. The disk density in the \texttt{relxill} model is fixed at $n_e\simeq10^{15}\:\mathrm{cm^{-3}}$ that can lead to a high value of iron abundance. So, we used \texttt{reflionx} based model to check the disk density. These models gave high disk density and reduced the iron abundance. The iron abundance came out to be $2.51_{-1.46}^{+1.48}$ using model M1.2 but the disk density was pegged at the upper bound. It is shown that the disk has high density. For the lamp-post geometry (Model M1.4), the iron abundance came out to be $2.12_{-0.50}^{+1.01}$. Our study shows that high iron abundance in XTE J2012+381 is due to high disk density.  
The \texttt{reflionx} based model clearly shows the high density for the disk but it is unable to determine the spin of BH. It only provided an upper limit, $a\lesssim0.7$, on spin value.

The final fit with \texttt{relxillCp/lpCp} (M1.3b and M1.5) are fitted for high density, and yet they still recover very high iron abundance. The results presented in table \ref{parameters} shows that the disk density in the \texttt{relxill} models has little effect in the recovered abundance of this source. While there appear to be important differences between the two reflection models in the high-density regime, their exact origin is still under study by the authors of these models, and thus the discrepancies in the model fit presented here could be taken as an indicative of the possible systematic uncertainties introduced by the reflection models (Garc\'ia, private communication).
 
The estimated total flux for two observations was found to be $6.74_{-0.01}^{+0.01}\times10^{-9}\;\mathrm{erg\;cm^{-2}\;s^{-1}}$ (Estimated by M1.5) and $8.84_{-0.01}^{+0.01}\times10^{-9}\;\mathrm{erg\;cm^{-2}\;s^{-1}}$ (Estimated by M2.2b) respectively. During the MJD 59942, the reflection component contributed $\sim21 \%$ and the disk contributed $\sim58 \%$ in the spectrum. During the MJD 59962, the disk contributed $\sim95\%$ in the spectrum. 

Thus, in conclusion, we studied the black hole candidate XTE J2012+381 during an outburst that started in late 2022 using the simultaneous data from \emph{Swift/XRT} and \emph{NuSTAR}. Two observations with a gap of 20 days are fitted and used to estimate the BH spin as well as the BH mass and inclination angle of the source. The distance of the source was taken as $5.4_{-1.50}^{+3.39}\;\mathrm{kpc}$. During MJD 59962, We used two models from the \texttt{relxill} family to estimate the spin of the black hole and inclination of the source, namely \texttt{relxillCp} with flatter inner emissivity profile and \texttt{relxilllpCp}. The two models estimate spin parameters of $0.883_{-0.061}^{+0.033}$ and $0.892_{-0.044}^{+0.020}$ and inclination angles of $46.2_{-2.0}^{+3.7}$ degrees and $43.1_{-1.2}^{+1.4}$ degrees, respectively. These values were further used for observation during MJD 59962  and estimated the mass of black hole $7.95_{-3.25}^{+7.65}\;\mathrm{M}_{\odot}$ and $7.48_{-2.75}^{+5.80}\;\mathrm{M}_{\odot}$, respectively.

\section*{Acknowledgements}
I am thankful to the referee for their insightful feedback and recommendations. I would like to thank Subir Bhattacharyya and Nilay Bhatt for their valuable opinions and discussions.
This research has made use of data and software obtained from the High Energy Astrophysics Science Archive Research Center (HEASARC), a service of the Astrophysics Science Division at NASA/GSFC and of the Smithsonian Astrophysical Observatory's High Energy Astrophysics Division.  We acknowledge the use of public data from the Swift data archive. This research has made use of the \emph{NuSTAR} Data Analysis Software (NuSTARDAS) jointly developed by the ASI Space Science Data Center (SSDC, Italy) and the California Institute of Technology (Caltech, USA).

\section*{Data Availability}

\noindent This  work  has  made  use  of  public  \emph{Swift/XRT}  data  available at 
\url{https://heasarc.gsfc.nasa.gov/FTP/swift/data/obs/}, \emph{NuSTAR} data available at 
\url{https://heasarc.gsfc.nasa.gov/FTP/nustar/data/obs/}.



\bibliographystyle{mnras}
\bibliography{MS_final} 

\begin{thebibliography}{}
\makeatletter
\relax
\def\mn@urlcharsother{\let\do\@makeother \do\$\do\&\do\#\do\^\do\_\do\%\do\~}
\def\mn@doi{\begingroup\mn@urlcharsother \@ifnextchar [ {\mn@doi@}
  {\mn@doi@[]}}
\def\mn@doi@[#1]#2{\def\@tempa{#1}\ifx\@tempa\@empty \href
  {http://dx.doi.org/#2} {doi:#2}\else \href {http://dx.doi.org/#2} {#1}\fi
  \endgroup}
\def\mn@eprint#1#2{\mn@eprint@#1:#2::\@nil}
\def\mn@eprint@arXiv#1{\href {http://arxiv.org/abs/#1} {{\tt arXiv:#1}}}
\def\mn@eprint@dblp#1{\href {http://dblp.uni-trier.de/rec/bibtex/#1.xml}
  {dblp:#1}}
\def\mn@eprint@#1:#2:#3:#4\@nil{\def\@tempa {#1}\def\@tempb {#2}\def\@tempc
  {#3}\ifx \@tempc \@empty \let \@tempc \@tempb \let \@tempb \@tempa \fi \ifx
  \@tempb \@empty \def\@tempb {arXiv}\fi \@ifundefined
  {mn@eprint@\@tempb}{\@tempb:\@tempc}{\expandafter \expandafter \csname
  mn@eprint@\@tempb\endcsname \expandafter{\@tempc}}}

\bibitem[\protect\citeauthoryear{{Abe}, {Fukazawa}, {Kubota}, {Kasama}  \&
  {Makishima}}{{Abe} et~al.}{2005}]{Abe2005}
{Abe} Y.,  {Fukazawa} Y.,  {Kubota} A.,  {Kasama} D.,   {Makishima} K.,  2005,
  \mn@doi [\pasj] {10.1093/pasj/57.4.629}, \href
  {https://ui.adsabs.harvard.edu/abs/2005PASJ...57..629A} {57, 629}

\bibitem[\protect\citeauthoryear{{Arnaud}}{{Arnaud}}{1996}]{Arnaud1996}
{Arnaud} K.~A.,  1996, in {Jacoby} G.~H.,  {Barnes} J.,  eds,  Astronomical
  Society of the Pacific Conference Series Vol. 101, Astronomical Data Analysis
  Software and Systems V. p.~17

\bibitem[\protect\citeauthoryear{{Burrows} et~al.,}{{Burrows}
  et~al.}{2005}]{Burrows2005}
{Burrows} D.~N.,  et~al., 2005, \mn@doi [\ssr] {10.1007/s11214-005-5097-2},
  \href {https://ui.adsabs.harvard.edu/abs/2005SSRv..120..165B} {120, 165}

\bibitem[\protect\citeauthoryear{{Campana}, {Stella}, {Belloni}, {Israel},
  {Santangelo}, {Frontera}, {Orlandini}  \& {Dal Fiume}}{{Campana}
  et~al.}{2002}]{Campana2002}
{Campana} S.,  {Stella} L.,  {Belloni} T.,  {Israel} G.~L.,  {Santangelo} A.,
  {Frontera} F.,  {Orlandini} M.,   {Dal Fiume} D.,  2002, \mn@doi [\aap]
  {10.1051/0004-6361:20020012}, \href
  {https://ui.adsabs.harvard.edu/abs/2002A&A...384..163C} {384, 163}

\bibitem[\protect\citeauthoryear{{Chakraborty}, {Ratheesh}, {Bhattacharyya},
  {Tomsick}, {Tombesi}, {Fukumura}  \& {Jaisawal}}{{Chakraborty}
  et~al.}{2021}]{Chakraborty2021}
{Chakraborty} S.,  {Ratheesh} A.,  {Bhattacharyya} S.,  {Tomsick} J.~A.,
  {Tombesi} F.,  {Fukumura} K.,   {Jaisawal} G.~K.,  2021, \mn@doi [\mnras]
  {10.1093/mnras/stab2530}, \href
  {https://ui.adsabs.harvard.edu/abs/2021MNRAS.508..475C} {508, 475}

\bibitem[\protect\citeauthoryear{{Dauser}, {Wilms}, {Reynolds}  \&
  {Brenneman}}{{Dauser} et~al.}{2010}]{Dauser2010}
{Dauser} T.,  {Wilms} J.,  {Reynolds} C.~S.,   {Brenneman} L.~W.,  2010,
  \mn@doi [\mnras] {10.1111/j.1365-2966.2010.17393.x}, \href
  {https://ui.adsabs.harvard.edu/abs/2010MNRAS.409.1534D} {409, 1534}

\bibitem[\protect\citeauthoryear{{Dauser}, {Garcia}, {Wilms}, {B{\"o}ck},
  {Brenneman}, {Falanga}, {Fukumura}  \& {Reynolds}}{{Dauser}
  et~al.}{2013}]{Dauser2013}
{Dauser} T.,  {Garcia} J.,  {Wilms} J.,  {B{\"o}ck} M.,  {Brenneman} L.~W.,
  {Falanga} M.,  {Fukumura} K.,   {Reynolds} C.~S.,  2013, \mn@doi [\mnras]
  {10.1093/mnras/sts710}, \href
  {https://ui.adsabs.harvard.edu/abs/2013MNRAS.430.1694D} {430, 1694}

\bibitem[\protect\citeauthoryear{{Dauser}, {Garcia}, {Parker}, {Fabian}  \&
  {Wilms}}{{Dauser} et~al.}{2014}]{Dauser2014}
{Dauser} T.,  {Garcia} J.,  {Parker} M.~L.,  {Fabian} A.~C.,   {Wilms} J.,
  2014, \mn@doi [\mnras] {10.1093/mnrasl/slu125}, \href
  {https://ui.adsabs.harvard.edu/abs/2014MNRAS.444L.100D} {444, L100}

\bibitem[\protect\citeauthoryear{{Davis} \& {El-Abd}}{{Davis} \&
  {El-Abd}}{2019}]{Davis2019}
{Davis} S.~W.,  {El-Abd} S.,  2019, \mn@doi [\apj] {10.3847/1538-4357/ab05c5},
  \href {https://ui.adsabs.harvard.edu/abs/2019ApJ...874...23D} {874, 23}

\bibitem[\protect\citeauthoryear{{Davis}, {Blaes}, {Hubeny}  \&
  {Turner}}{{Davis} et~al.}{2005}]{Davis2005}
{Davis} S.~W.,  {Blaes} O.~M.,  {Hubeny} I.,   {Turner} N.~J.,  2005, \mn@doi
  [\apj] {10.1086/427278}, \href
  {https://ui.adsabs.harvard.edu/abs/2005ApJ...621..372D} {621, 372}

\bibitem[\protect\citeauthoryear{{Dong}, {Garc{\'\i}a}, {Liu}, {Zhao}, {Zheng}
  \& {Gou}}{{Dong} et~al.}{2020a}]{Dong2020_1836}
{Dong} Y.,  {Garc{\'\i}a} J.~A.,  {Liu} Z.,  {Zhao} X.,  {Zheng} X.,   {Gou}
  L.,  2020a, \mn@doi [\mnras] {10.1093/mnras/staa401}, \href
  {https://ui.adsabs.harvard.edu/abs/2020MNRAS.493.2178D} {493, 2178}

\bibitem[\protect\citeauthoryear{{Dong}, {Garc{\'\i}a}, {Steiner}  \&
  {Gou}}{{Dong} et~al.}{2020b}]{Dong2020_1543}
{Dong} Y.,  {Garc{\'\i}a} J.~A.,  {Steiner} J.~F.,   {Gou} L.,  2020b, \mn@doi
  [\mnras] {10.1093/mnras/staa606}, \href
  {https://ui.adsabs.harvard.edu/abs/2020MNRAS.493.4409D} {493, 4409}

\bibitem[\protect\citeauthoryear{{Draghis}, {Miller}, {Brumback}, {Fabian},
  {Tomsick}  \& {Zoghbi}}{{Draghis} et~al.}{2023a}]{Draghis2023}
{Draghis} P.~A.,  {Miller} J.~M.,  {Brumback} M.~C.,  {Fabian} A.~C.,
  {Tomsick} J.~A.,   {Zoghbi} A.,  2023a, \mn@doi [arXiv e-prints]
  {10.48550/arXiv.2307.06988}, \href
  {https://ui.adsabs.harvard.edu/abs/2023arXiv230706988D} {p. arXiv:2307.06988}

\bibitem[\protect\citeauthoryear{{Draghis}, {Miller}, {Zoghbi}, {Reynolds},
  {Costantini}, {Gallo}  \& {Tomsick}}{{Draghis} et~al.}{2023b}]{Draghis2023a}
{Draghis} P.~A.,  {Miller} J.~M.,  {Zoghbi} A.,  {Reynolds} M.,  {Costantini}
  E.,  {Gallo} L.~C.,   {Tomsick} J.~A.,  2023b, \mn@doi [\apj]
  {10.3847/1538-4357/acafe7}, \href
  {https://ui.adsabs.harvard.edu/abs/2023ApJ...946...19D} {946, 19}

\bibitem[\protect\citeauthoryear{{Dunn}, {Fender}, {K{\"o}rding}, {Belloni}  \&
  {Cabanac}}{{Dunn} et~al.}{2010}]{Dunn2010}
{Dunn} R.~J.~H.,  {Fender} R.~P.,  {K{\"o}rding} E.~G.,  {Belloni} T.,
  {Cabanac} C.,  2010, \mn@doi [\mnras] {10.1111/j.1365-2966.2010.16114.x},
  \href {https://ui.adsabs.harvard.edu/abs/2010MNRAS.403...61D} {403, 61}

\bibitem[\protect\citeauthoryear{{Fabian}, {Rees}, {Stella}  \&
  {White}}{{Fabian} et~al.}{1989}]{Febian1989}
{Fabian} A.~C.,  {Rees} M.~J.,  {Stella} L.,   {White} N.~E.,  1989, \mn@doi
  [\mnras] {10.1093/mnras/238.3.729}, \href
  {https://ui.adsabs.harvard.edu/abs/1989MNRAS.238..729F} {238, 729}

\bibitem[\protect\citeauthoryear{{Fabian}, {Iwasawa}, {Reynolds}  \&
  {Young}}{{Fabian} et~al.}{2000}]{Febian2000}
{Fabian} A.~C.,  {Iwasawa} K.,  {Reynolds} C.~S.,   {Young} A.~J.,  2000,
  \mn@doi [\pasp] {10.1086/316610}, \href
  {https://ui.adsabs.harvard.edu/abs/2000PASP..112.1145F} {112, 1145}

\bibitem[\protect\citeauthoryear{{Fabian} et~al.,}{{Fabian}
  et~al.}{2012}]{Fabian2012}
{Fabian} A.~C.,  et~al., 2012, \mn@doi [\mnras]
  {10.1111/j.1365-2966.2012.21185.x}, \href
  {https://ui.adsabs.harvard.edu/abs/2012MNRAS.424..217F} {424, 217}

\bibitem[\protect\citeauthoryear{{Gaia Collaboration} et~al.,}{{Gaia
  Collaboration} et~al.}{2016}]{Gaia2016}
{Gaia Collaboration} et~al., 2016, \mn@doi [\aap]
  {10.1051/0004-6361/201629272}, \href
  {https://ui.adsabs.harvard.edu/abs/2016A&A...595A...1G} {595, A1}

\bibitem[\protect\citeauthoryear{{Garc{\'\i}a} \& {Kallman}}{{Garc{\'\i}a} \&
  {Kallman}}{2010}]{Garcia2010}
{Garc{\'\i}a} J.,  {Kallman} T.~R.,  2010, \mn@doi [\apj]
  {10.1088/0004-637X/718/2/695}, \href
  {https://ui.adsabs.harvard.edu/abs/2010ApJ...718..695G} {718, 695}

\bibitem[\protect\citeauthoryear{{Garc{\'\i}a}, {Kallman}  \&
  {Mushotzky}}{{Garc{\'\i}a} et~al.}{2011}]{Garcia2011}
{Garc{\'\i}a} J.,  {Kallman} T.~R.,   {Mushotzky} R.~F.,  2011, \mn@doi [\apj]
  {10.1088/0004-637X/731/2/131}, \href
  {https://ui.adsabs.harvard.edu/abs/2011ApJ...731..131G} {731, 131}

\bibitem[\protect\citeauthoryear{{Garc{\'\i}a}, {Dauser}, {Reynolds},
  {Kallman}, {McClintock}, {Wilms}  \& {Eikmann}}{{Garc{\'\i}a}
  et~al.}{2013}]{Garcia2013}
{Garc{\'\i}a} J.,  {Dauser} T.,  {Reynolds} C.~S.,  {Kallman} T.~R.,
  {McClintock} J.~E.,  {Wilms} J.,   {Eikmann} W.,  2013, \mn@doi [\apj]
  {10.1088/0004-637X/768/2/146}, \href
  {https://ui.adsabs.harvard.edu/abs/2013ApJ...768..146G} {768, 146}

\bibitem[\protect\citeauthoryear{{Garc{\'\i}a} et~al.,}{{Garc{\'\i}a}
  et~al.}{2014}]{Garcia2014}
{Garc{\'\i}a} J.,  et~al., 2014, \mn@doi [\apj] {10.1088/0004-637X/782/2/76},
  \href {https://ui.adsabs.harvard.edu/abs/2014ApJ...782...76G} {782, 76}

\bibitem[\protect\citeauthoryear{{Garc{\'\i}a}, {Steiner}, {McClintock},
  {Remillard}, {Grinberg}  \& {Dauser}}{{Garc{\'\i}a}
  et~al.}{2015}]{Garcia2015}
{Garc{\'\i}a} J.~A.,  {Steiner} J.~F.,  {McClintock} J.~E.,  {Remillard} R.~A.,
   {Grinberg} V.,   {Dauser} T.,  2015, \mn@doi [\apj]
  {10.1088/0004-637X/813/2/84}, \href
  {https://ui.adsabs.harvard.edu/abs/2015ApJ...813...84G} {813, 84}

\bibitem[\protect\citeauthoryear{{Garc{\'\i}a}, {Kallman}, {Bautista},
  {Mendoza}, {Deprince}, {Palmeri}  \& {Quinet}}{{Garc{\'\i}a}
  et~al.}{2018}]{Garcia2018}
{Garc{\'\i}a} J.~A.,  {Kallman} T.~R.,  {Bautista} M.,  {Mendoza} C.,
  {Deprince} J.,  {Palmeri} P.,   {Quinet} P.,  2018, in Workshop on
  Astrophysical Opacities. p.~282 (\mn@eprint {arXiv} {1805.00581}),
  \mn@doi{10.48550/arXiv.1805.00581}

\bibitem[\protect\citeauthoryear{{Gehrels} et~al.,}{{Gehrels}
  et~al.}{2004}]{Gehrels2004}
{Gehrels} N.,  et~al., 2004, \mn@doi [\apj] {10.1086/422091}, \href
  {https://ui.adsabs.harvard.edu/abs/2004ApJ...611.1005G} {611, 1005}

\bibitem[\protect\citeauthoryear{{Gierli{\'n}ski} \& {Done}}{{Gierli{\'n}ski}
  \& {Done}}{2004}]{Gierlinski2004}
{Gierli{\'n}ski} M.,  {Done} C.,  2004, \mn@doi [\mnras]
  {10.1111/j.1365-2966.2004.07266.x}, \href
  {https://ui.adsabs.harvard.edu/abs/2004MNRAS.347..885G} {347, 885}

\bibitem[\protect\citeauthoryear{{Harrison} et~al.,}{{Harrison}
  et~al.}{2013}]{Harrison2013}
{Harrison} F.~A.,  et~al., 2013, \mn@doi [\apj] {10.1088/0004-637X/770/2/103},
  \href {https://ui.adsabs.harvard.edu/abs/2013ApJ...770..103H} {770, 103}

\bibitem[\protect\citeauthoryear{{Hynes}, {Roche}, {Charles}  \& {Coe}}{{Hynes}
  et~al.}{1999}]{Hynes1999}
{Hynes} R.~I.,  {Roche} P.,  {Charles} P.~A.,   {Coe} M.~J.,  1999, \mn@doi
  [\mnras] {10.1046/j.1365-8711.1999.02653.x}, \href
  {https://ui.adsabs.harvard.edu/abs/1999MNRAS.305L..49H} {305, L49}

\bibitem[\protect\citeauthoryear{{Jana}, {Naik}, {Chatterjee}  \&
  {Jaisawal}}{{Jana} et~al.}{2021}]{Jana2021}
{Jana} A.,  {Naik} S.,  {Chatterjee} D.,   {Jaisawal} G.~K.,  2021, \mn@doi
  [\mnras] {10.1093/mnras/stab2448}, \href
  {https://ui.adsabs.harvard.edu/abs/2021MNRAS.507.4779J} {507, 4779}

\bibitem[\protect\citeauthoryear{{Jia}, {Feng}, {Song}, {Yang}, {Yuh}, {Huang}
  \& {Gou}}{{Jia} et~al.}{2023}]{Jia2023}
{Jia} N.,  {Feng} Y.,  {Song} Y.-J.,  {Yang} J.,  {Yuh} J.,  {Huang} P.-J.,
  {Gou} L.-J.,  2023, \mn@doi [Research in Astronomy and Astrophysics]
  {10.1088/1674-4527/acd58c}, \href
  {https://ui.adsabs.harvard.edu/abs/2023RAA....23g5022J} {23, 075022}

\bibitem[\protect\citeauthoryear{{Kawamuro} et~al.,}{{Kawamuro}
  et~al.}{2022}]{Kawamuro2022}
{Kawamuro} T.,  et~al., 2022, The Astronomer's Telegram, \href
  {https://ui.adsabs.harvard.edu/abs/2022ATel15826....1K} {15826, 1}

\bibitem[\protect\citeauthoryear{{Kennea}}{{Kennea}}{2022}]{Kennea2022}
{Kennea} J.~A.,  2022, The Astronomer's Telegram, \href
  {https://ui.adsabs.harvard.edu/abs/2022ATel15827....1K} {15827, 1}

\bibitem[\protect\citeauthoryear{{Kumar}, {Bhattacharyya}, {Bhatt}  \&
  {Misra}}{{Kumar} et~al.}{2022}]{Kumar2022}
{Kumar} R.,  {Bhattacharyya} S.,  {Bhatt} N.,   {Misra} R.,  2022, \mn@doi
  [\mnras] {10.1093/mnras/stac1170}, \href
  {https://ui.adsabs.harvard.edu/abs/2022MNRAS.513.4869K} {513, 4869}

\bibitem[\protect\citeauthoryear{{Li}, {Zimmerman}, {Narayan}  \&
  {McClintock}}{{Li} et~al.}{2005}]{Li2005}
{Li} L.-X.,  {Zimmerman} E.~R.,  {Narayan} R.,   {McClintock} J.~E.,  2005,
  \mn@doi [\apjs] {10.1086/428089}, \href
  {https://ui.adsabs.harvard.edu/abs/2005ApJS..157..335L} {157, 335}

\bibitem[\protect\citeauthoryear{{Liu} et~al.,}{{Liu} et~al.}{2023}]{Liu2023}
{Liu} H.,  et~al., 2023, \mn@doi [\apj] {10.3847/1538-4357/acd8b9}, \href
  {https://ui.adsabs.harvard.edu/abs/2023ApJ...951..145L} {951, 145}

\bibitem[\protect\citeauthoryear{{Makishima}, {Maejima}, {Mitsuda}, {Bradt},
  {Remillard}, {Tuohy}, {Hoshi}  \& {Nakagawa}}{{Makishima}
  et~al.}{1986}]{Makishima1986}
{Makishima} K.,  {Maejima} Y.,  {Mitsuda} K.,  {Bradt} H.~V.,  {Remillard}
  R.~A.,  {Tuohy} I.~R.,  {Hoshi} R.,   {Nakagawa} M.,  1986, \mn@doi [\apj]
  {10.1086/164534}, \href
  {https://ui.adsabs.harvard.edu/abs/1986ApJ...308..635M} {308, 635}

\bibitem[\protect\citeauthoryear{{McClintock}, {Remillard}, {Rupen}, {Torres},
  {Steeghs}, {Levine}  \& {Orosz}}{{McClintock} et~al.}{2009}]{McClintock2009}
{McClintock} J.~E.,  {Remillard} R.~A.,  {Rupen} M.~P.,  {Torres} M.~A.~P.,
  {Steeghs} D.,  {Levine} A.~M.,   {Orosz} J.~A.,  2009, \mn@doi [\apj]
  {10.1088/0004-637X/698/2/1398}, \href
  {https://ui.adsabs.harvard.edu/abs/2009ApJ...698.1398M} {698, 1398}

\bibitem[\protect\citeauthoryear{{McClintock}, {Narayan}  \&
  {Steiner}}{{McClintock} et~al.}{2014}]{McClintock2014}
{McClintock} J.~E.,  {Narayan} R.,   {Steiner} J.~F.,  2014, \mn@doi [\ssr]
  {10.1007/s11214-013-0003-9}, \href
  {https://ui.adsabs.harvard.edu/abs/2014SSRv..183..295M} {183, 295}

\bibitem[\protect\citeauthoryear{{Miller}}{{Miller}}{2007}]{Miller2007}
{Miller} J.~M.,  2007, \mn@doi [\araa]
  {10.1146/annurev.astro.45.051806.110555}, \href
  {https://ui.adsabs.harvard.edu/abs/2007ARA&A..45..441M} {45, 441}

\bibitem[\protect\citeauthoryear{{Mitsuda} et~al.,}{{Mitsuda}
  et~al.}{1984}]{Mitsuda1984}
{Mitsuda} K.,  et~al., 1984, \pasj, \href
  {https://ui.adsabs.harvard.edu/abs/1984PASJ...36..741M} {36, 741}

\bibitem[\protect\citeauthoryear{{Noble}, {Krolik}  \& {Hawley}}{{Noble}
  et~al.}{2010}]{Noble2010}
{Noble} S.~C.,  {Krolik} J.~H.,   {Hawley} J.~F.,  2010, \mn@doi [\apj]
  {10.1088/0004-637X/711/2/959}, \href
  {https://ui.adsabs.harvard.edu/abs/2010ApJ...711..959N} {711, 959}

\bibitem[\protect\citeauthoryear{{Novikov} \& {Thorne}}{{Novikov} \&
  {Thorne}}{1973}]{NT1973}
{Novikov} I.~D.,  {Thorne} K.~S.,  1973, in Black Holes (Les Astres Occlus). pp
  343--450

\bibitem[\protect\citeauthoryear{{Parker} et~al.,}{{Parker}
  et~al.}{2015}]{Parker2015}
{Parker} M.~L.,  et~al., 2015, \mn@doi [\apj] {10.1088/0004-637X/808/1/9},
  \href {https://ui.adsabs.harvard.edu/abs/2015ApJ...808....9P} {808, 9}

\bibitem[\protect\citeauthoryear{{Parker} et~al.,}{{Parker}
  et~al.}{2016}]{Parker2016}
{Parker} M.~L.,  et~al., 2016, \mn@doi [\apjl] {10.3847/2041-8205/821/1/L6},
  \href {https://ui.adsabs.harvard.edu/abs/2016ApJ...821L...6P} {821, L6}

\bibitem[\protect\citeauthoryear{{Penna}, {McKinney}, {Narayan},
  {Tchekhovskoy}, {Shafee}  \& {McClintock}}{{Penna} et~al.}{2010}]{Penna2010}
{Penna} R.~F.,  {McKinney} J.~C.,  {Narayan} R.,  {Tchekhovskoy} A.,  {Shafee}
  R.,   {McClintock} J.~E.,  2010, \mn@doi [\mnras]
  {10.1111/j.1365-2966.2010.17170.x}, \href
  {https://ui.adsabs.harvard.edu/abs/2010MNRAS.408..752P} {408, 752}

\bibitem[\protect\citeauthoryear{{Prabhakar}, {Mandal}, {Bhuvana}  \&
  {Nandi}}{{Prabhakar} et~al.}{2023}]{Prabhakar2023}
{Prabhakar} G.,  {Mandal} S.,  {Bhuvana} G.~R.,   {Nandi} A.,  2023, \mn@doi
  [\mnras] {10.1093/mnras/stad080}, \href
  {https://ui.adsabs.harvard.edu/abs/2023MNRAS.520.4889P} {520, 4889}

\bibitem[\protect\citeauthoryear{{Remillard} et~al.,}{{Remillard}
  et~al.}{1998}]{Remillard1998}
{Remillard} R.,  et~al., 1998, \iaucirc, \href
  {https://ui.adsabs.harvard.edu/abs/1998IAUC.6920....1R} {6920, 1}

\bibitem[\protect\citeauthoryear{{Reynolds}}{{Reynolds}}{2014}]{Reynolds2014}
{Reynolds} C.~S.,  2014, \mn@doi [\ssr] {10.1007/s11214-013-0006-6}, \href
  {https://ui.adsabs.harvard.edu/abs/2014SSRv..183..277R} {183, 277}

\bibitem[\protect\citeauthoryear{{Reynolds}}{{Reynolds}}{2021}]{Reynolds2021}
{Reynolds} C.~S.,  2021, \mn@doi [\araa] {10.1146/annurev-astro-112420-035022},
  \href {https://ui.adsabs.harvard.edu/abs/2021ARA&A..59..117R} {59, 117}

\bibitem[\protect\citeauthoryear{{Reynolds} \& {Fabian}}{{Reynolds} \&
  {Fabian}}{2008}]{Reynolds2008}
{Reynolds} C.~S.,  {Fabian} A.~C.,  2008, \mn@doi [\apj] {10.1086/527344},
  \href {https://ui.adsabs.harvard.edu/abs/2008ApJ...675.1048R} {675, 1048}

\bibitem[\protect\citeauthoryear{{Reynolds} \& {Nowak}}{{Reynolds} \&
  {Nowak}}{2003}]{Reynolds2003}
{Reynolds} C.~S.,  {Nowak} M.~A.,  2003, \mn@doi [\physrep]
  {10.1016/S0370-1573(02)00584-7}, \href
  {https://ui.adsabs.harvard.edu/abs/2003PhR...377..389R} {377, 389}

\bibitem[\protect\citeauthoryear{{Romano} et~al.,}{{Romano}
  et~al.}{2006}]{Romano2006}
{Romano} P.,  et~al., 2006, \mn@doi [\aap] {10.1051/0004-6361:20065071}, \href
  {https://ui.adsabs.harvard.edu/abs/2006A&A...456..917R} {456, 917}

\bibitem[\protect\citeauthoryear{{Ross} \& {Fabian}}{{Ross} \&
  {Fabian}}{2005}]{Ross2005}
{Ross} R.~R.,  {Fabian} A.~C.,  2005, \mn@doi [\mnras]
  {10.1111/j.1365-2966.2005.08797.x}, \href
  {https://ui.adsabs.harvard.edu/abs/2005MNRAS.358..211R} {358, 211}

\bibitem[\protect\citeauthoryear{{Schnittman}, {Krolik}  \&
  {Noble}}{{Schnittman} et~al.}{2013}]{Schnittman2013}
{Schnittman} J.~D.,  {Krolik} J.~H.,   {Noble} S.~C.,  2013, \mn@doi [\apj]
  {10.1088/0004-637X/769/2/156}, \href
  {https://ui.adsabs.harvard.edu/abs/2013ApJ...769..156S} {769, 156}

\bibitem[\protect\citeauthoryear{{Shafee}, {Narayan}  \& {McClintock}}{{Shafee}
  et~al.}{2008}]{Shafee2008}
{Shafee} R.,  {Narayan} R.,   {McClintock} J.~E.,  2008, \mn@doi [\apj]
  {10.1086/527346}, \href
  {https://ui.adsabs.harvard.edu/abs/2008ApJ...676..549S} {676, 549}

\bibitem[\protect\citeauthoryear{{Shakura} \& {Sunyaev}}{{Shakura} \&
  {Sunyaev}}{1973}]{SS1973}
{Shakura} N.~I.,  {Sunyaev} R.~A.,  1973, \aap, \href
  {https://ui.adsabs.harvard.edu/abs/1973A&A....24..337S} {24, 337}

\bibitem[\protect\citeauthoryear{{Shimura} \& {Takahara}}{{Shimura} \&
  {Takahara}}{1995}]{Shimura1995}
{Shimura} T.,  {Takahara} F.,  1995, \mn@doi [\apj] {10.1086/175740}, \href
  {https://ui.adsabs.harvard.edu/abs/1995ApJ...445..780S} {445, 780}

\bibitem[\protect\citeauthoryear{{Tanaka} et~al.,}{{Tanaka}
  et~al.}{1995}]{Tanaka1995}
{Tanaka} Y.,  et~al., 1995, \mn@doi [\nat] {10.1038/375659a0}, \href
  {https://ui.adsabs.harvard.edu/abs/1995Natur.375..659T} {375, 659}

\bibitem[\protect\citeauthoryear{{Tao}, {Tomsick}, {Qu}, {Zhang}, {Zhang}  \&
  {Bu}}{{Tao} et~al.}{2019}]{Tao2019}
{Tao} L.,  {Tomsick} J.~A.,  {Qu} J.,  {Zhang} S.,  {Zhang} S.,   {Bu} Q.,
  2019, \mn@doi [\apj] {10.3847/1538-4357/ab5282}, \href
  {https://ui.adsabs.harvard.edu/abs/2019ApJ...887..184T} {887, 184}

\bibitem[\protect\citeauthoryear{{Walton} et~al.,}{{Walton}
  et~al.}{2016}]{Walton2016}
{Walton} D.~J.,  et~al., 2016, \mn@doi [\apj] {10.3847/0004-637X/826/1/87},
  \href {https://ui.adsabs.harvard.edu/abs/2016ApJ...826...87W} {826, 87}

\bibitem[\protect\citeauthoryear{{Walton} et~al.,}{{Walton}
  et~al.}{2017}]{Walton2017}
{Walton} D.~J.,  et~al., 2017, \mn@doi [\apj] {10.3847/1538-4357/aa67e8}, \href
  {https://ui.adsabs.harvard.edu/abs/2017ApJ...839..110W} {839, 110}

\bibitem[\protect\citeauthoryear{{White}, {Ueda}, {Dotani}  \&
  {Nagase}}{{White} et~al.}{1998}]{White1998}
{White} N.~E.,  {Ueda} Y.,  {Dotani} T.,   {Nagase} F.,  1998, \iaucirc, \href
  {https://ui.adsabs.harvard.edu/abs/1998IAUC.6927....2W} {6927, 2}

\bibitem[\protect\citeauthoryear{{Wilkins} \& {Fabian}}{{Wilkins} \&
  {Fabian}}{2012}]{Wilkins2012}
{Wilkins} D.~R.,  {Fabian} A.~C.,  2012, \mn@doi [\mnras]
  {10.1111/j.1365-2966.2012.21308.x}, \href
  {https://ui.adsabs.harvard.edu/abs/2012MNRAS.424.1284W} {424, 1284}

\bibitem[\protect\citeauthoryear{{Wilms}, {Allen}  \& {McCray}}{{Wilms}
  et~al.}{2000}]{Wilms2000}
{Wilms} J.,  {Allen} A.,   {McCray} R.,  2000, \mn@doi [\apj] {10.1086/317016},
  \href {https://ui.adsabs.harvard.edu/abs/2000ApJ...542..914W} {542, 914}

\bibitem[\protect\citeauthoryear{{Zdziarski}, {Johnson}  \&
  {Magdziarz}}{{Zdziarski} et~al.}{1996}]{Zdziarski1996}
{Zdziarski} A.~A.,  {Johnson} W.~N.,   {Magdziarz} P.,  1996, \mn@doi [\mnras]
  {10.1093/mnras/283.1.193}, \href
  {https://ui.adsabs.harvard.edu/abs/1996MNRAS.283..193Z} {283, 193}

\bibitem[\protect\citeauthoryear{{Zhang}, {Cui}  \& {Chen}}{{Zhang}
  et~al.}{1997}]{Zhang1997}
{Zhang} S.~N.,  {Cui} W.,   {Chen} W.,  1997, \mn@doi [\apjl] {10.1086/310705},
  \href {https://ui.adsabs.harvard.edu/abs/1997ApJ...482L.155Z} {482, L155}

\bibitem[\protect\citeauthoryear{{{\.Z}ycki}, {Done}  \& {Smith}}{{{\.Z}ycki}
  et~al.}{1999}]{Zycki1999}
{{\.Z}ycki} P.~T.,  {Done} C.,   {Smith} D.~A.,  1999, \mn@doi [\mnras]
  {10.1046/j.1365-8711.1999.02885.x}, \href
  {https://ui.adsabs.harvard.edu/abs/1999MNRAS.309..561Z} {309, 561}

\makeatother
\end{thebibliography}








\bsp	
\label{lastpage}
\end{document}